\documentclass[12pt, a4paper,superscriptaddress,groupedaddress]{revtex4}   
\usepackage{amsmath}
\usepackage{bbm}
\usepackage{graphicx}  
\usepackage{dcolumn}  
\usepackage{bm}        
\usepackage{amssymb}

\begin{document}

\title{Verbal Perception and the Word Length Effect}

\author{Francesco Fumarola}
\noaffiliation

\date{\today}

\begin{abstract}   
A theoretical framework is proposed for the understanding of verbal perception -- the conversion of words into meaning, modeled as a compromise between lexical demands and contextual constraints -- and the theory is tested against experiments on short-term memory.  The observation that lists of short words are recalled better than lists of long ones has been a long-standing subject of controversy, further complicated by the apparent inversion of the effect for mixed lists. In the framework here proposed, these behaviors emerge as an effect of the different level of localization of short and long words in semantic space. Events corresponding to the recognition of a nonlocal word have a clustering property in phase space, which facilitates associative retrieval. The standard word-length effect arises directly from this property, and the inverse effect from its breakdown. An analysis of data from the PEERS experiments (Healey and Kahana, 2016) confirms the main predictions of the theory. Further predictions are listed and new experiments are proposed. Finally, an interpretation of the above results is presented.
\end{abstract}    
         
\maketitle

\section{Introduction}

\subsection{Free recall: lexical and serial-position effects }

In 1894, with their pioneering work on free-recall experiments, Binet and Henry introduced a key tool for the controlled investigation of short-term memory (Binet and Henry, 1894). In its traditional form, a free-recall experiment is performed by presenting the subject with a list of words and then requesting him or her to recall it in any order (Murdock, 1960, 1962; Roberts, 1972; Standing, 1973).

Several types of effects have been reported: 

1) Effects depending on the lexical properties of individual words. In particular, lists of short words are recalled better than lists of long ones, a fact known in the literature as the word-length effect (Baddeley et al., 1975; Russo and Grammatopoulou, 2003; Tehan and Tolan, 2007; Bhatarah et al., 2009). 

2) Effects in which the recall probability depends on the absolute position of words in the list. It has been observed that the first and last words in the list are recalled more easily ("primacy" and "recency" effects). 

3) Effects depending on the relative position of words with respect to each other. Most notably, the recall probabilities of contiguous words correlate positively, a fact known as the contiguity effect (Murdock, 1960, 1962). 

The need to understand serial-position effects led to the devising of retrieved-context models, such as the Temporal Context Model (Howard and Kanaha, 2002). In these models the recall process, rather than retrieving a word directly, retrieves the context associated to the word first. 

Within this scenario, recency effects appear because the context at the time of the "memory test" is most similar to the context associated with recent items. When an item is retrieved, it reinstates the context active when that item was presented. Because this context overlaps with the encoding context of the items' neighbors, a contiguity effect results. 

Through these models, serial-position effects have been substantially understood over the past fifteen years (Howard and Kahana, 2002, 2002b; Sederberg, Howard, and Kahana, 2008; Polyn, Norman, and Kahana,  2009, 2009b;  Lohnas, Polyn, and Kahana, 2015;  Kahana, 2012). The same cannot be said, however, about the word-length effect. 
 
\subsection{Riddles of the word length effect}

The word-length effect (WLE) has been a traditional testing ground for models of short-term memory (Campoy, 2011; Jalbert et al., 2011), and it has played a key role in establishing the working-memory paradigm and the phonological loop hypothesis (Baddeley and Hitch., 1974). 
 
The standard account of the effect (Baddeley, 2007) relies on a trade-off between memory decay (in the phonological store) and subvocal rehearsal via an articulatory control process. Because shorter words take less time to rehearse, more decaying traces of them can be refreshed than decaying traces of long items, and, therefore, more short items can be recalled. 

This picture, however, is not able to account for all experimental observations concerning this effect, and has been repeatedly called into question.  

In (Neath et al., 2003), it was shown that with words having the same number of syllables but different pronunciation times, no unambiguous WLE arises. This result (extended in Jalbert et al. 2011) suggests that the effect depends on the number of syllables, and not on the time it takes to pronounce them.

Experiments have also been performed in conditions where there was a delay between lists, making subvocal rehearsal possible in the interval. No appreciable difference in recall probabilities was found (Campoy, 2008). 

In the same study, experiments were performed in which subvocal rehearsal was prevented by a high presentation rate. No delay was allowed between the presentation of word lists and the memory test. Yet, the WLE occurred unperturbed. 

In the 2011 paper I just cited,  Jalbert et al. concluded: "the WLE may be better explained by the 
differences in linguistic and lexical properties of short and long words rather than by length per se".

\subsection{Semantic predictors of word length }
  
In the meanwhile, within the fields of experimental and computational linguistics, progress has been made in understanding the role of word length in verbal processing. Over the years, it has emerged that words with different lengths tend to have different semantic properties. 

The idea was first put forth in pedagogical studies  
(Klare, 1988). In (Elts, 1995), a correlation coefficient of 0.96 was found between a noun's length and its average tendency to be used as a technical term ("terminologicality"). Mikk et al. (2000), using data on the human-assessed complexity of a large sample of words, found a correlation coefficient 0.86 between words' length and their semantic complexity.  
     
Pinning down the precise semantic property that correlates to word length has proven difficult. Already in (Greenberg, 1966) it was argued that a word's length correlates positively to its 
conceptual "markedness" of meaning. Various notions of markedness have subsequently been discussed in the literature (Haspelmath, 2006). 
    
Piantadosi et al. (Piantadosi et al., 2011, 2011B) and later Mahowald et al., (Mahowald et al., 2012) reported that the length of words correlates positively with their contextual information rate. More recently, Lewis and Frank (2016) have carried out a comprehensive experimental study across 80 languages. They found that, in all the languages considered, judgments of conceptual complexity for a sample of real words correlate highly with their length, and they even control
for frequency, familiarity, imageability, and concreteness. Their conclusion is: "While word lengths are systematically
related to usage $\--$ both frequency and contextual predictability $\--$ our results reveal a systematic
relationship with meaning as well". 
    
In the light of these findings, it would be a natural step to attempt an explanation of the WLE in terms of the semantic differences among words. However, no such approach seems to have been attempted in the literature. 

\subsection{The inverse word length effect}

Recently, new aspects of the WLE have emerged through the analysis of a large set of data from experiments by Miller and al (Miller et al., 2012). The data analysis was performed by Katkov et al. (Katkov et al., 2014), who found no negative correlation between total length of presented items and number of recalled words, thus disproving both rehearsal-time theories and hypotheses based on the increasing complexity of longer items. 

Moreover, they reported an inversion of the effect in mixed lists, that is, lists where words are selected irrespectively of their length. They observed that, in this type of lists, the mean values of recall probabilities allow to establish an increasing trend. Long words are recalled better than short ones.

An "inverse" WLE had been previously reported by at least two groups, but in somewhat less general circumstances: one of them (Hulme et al., 2006) embedded strictly pure lists with a single word of a different type, while the results of Xu et al. (Xu et al., 2009), may not bear direct comparison with data in languages other than Chinese.
 
If the inversion of the WLE for mixed lists will be confirmed by further experiments, it will have to be taken into account by every general theory of the standard WLE. Let us consider, therefore, what requirements a model should fulfill to explain both phenomena simultaneously.
 
Call $\gamma$ the fraction of long words in the list; call $P_l(\gamma)$ the probability of recalling successfully a given long word from a list in which a fraction $\gamma$ of words are long; and let $P_s(\gamma)$ be the probability of recalling successfully a short word, from a list with a fraction $\gamma$ of long words. 

Obviously, the function $P_l(\gamma)$ is only defined for $\gamma>0$, and the function $P_s(\gamma)$ only for $\gamma<1$. For $\gamma \in ]0, 1[$, both functions are defined.

Theorists would have to reconcile two observations on the curves $P_s(\gamma), P_l(\gamma)$:

1. $P_l(\gamma=1) < P_s(\gamma=0)$
 
2. $P_l(\gamma) > P_s(\gamma)$  for $\gamma \in ]0, 1[$.

The only way these two inequalities can be simultaneously satisfied is if \textit{both} $P_l$ and $P_s$ are, on the whole, decreasing functions of $\gamma$. The simplest choice of these curves compatible with experiments is one where both are monotonously decreasing, that is:
 
\begin{equation}
\label{derivatives}
\frac{d P_l}{d \gamma } < 0 \ \ \ \ \ \frac{d P_s}{d \gamma} < 0
\end{equation}

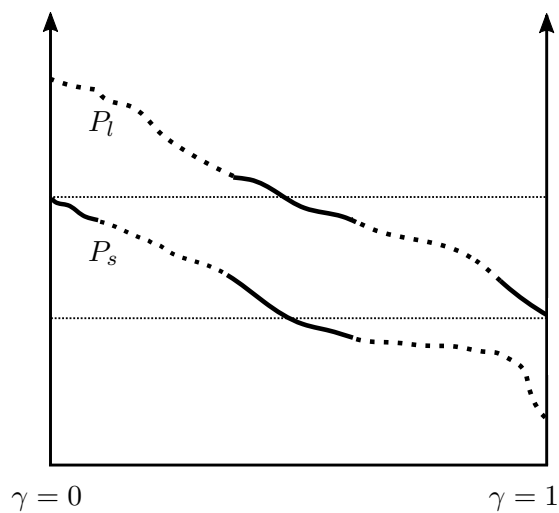
\begin{figure}[h]
\label{fig1}
\includegraphics[width=0.5\textwidth]{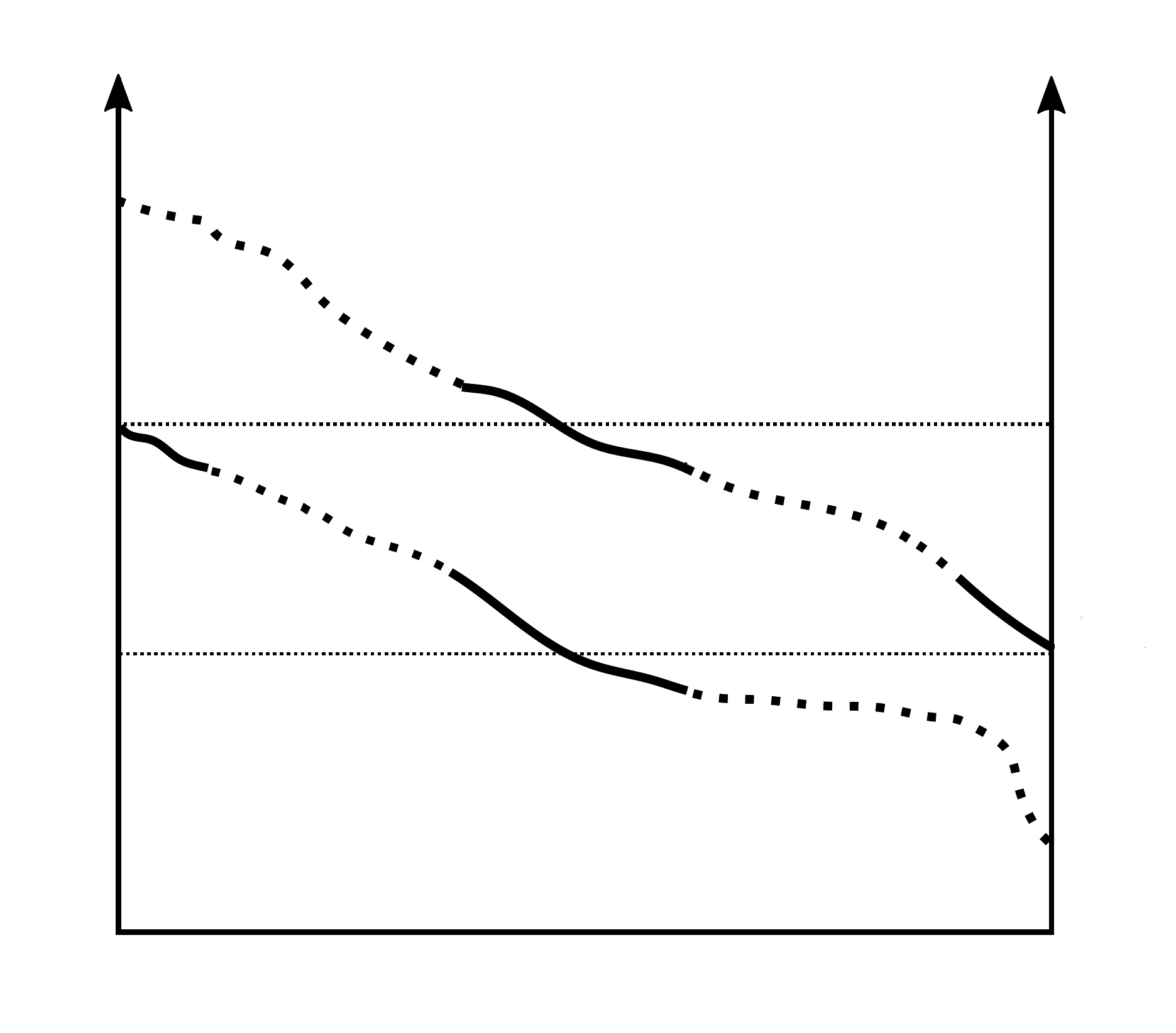}
\put(-225,5){$\gamma=0$}
\put(-50,5){ $\gamma= 1$}
\put(-200,146){ $P_l$}
\put(-200,97){ $P_s$}
\caption{Back-of-envelope sketch of the possible behavior of recall probabilities for short and long words, as a function of the fraction of long words in the list.}
\centering
\end{figure}

This means that, whenever we replace a short word of the list with a long word, we are lowering the recall probability of all the words in the list, both long and short. A higher number of long words makes every single word in the list harder to recall.

It is difficult to imagine how this could ensue from the different $\textit{duration}$ of long and short words. The question is then: can equation (\ref{derivatives}) result directly from the different semantic properties of long and short words? 

In this paper, I will show that the answer is positive as long as one models carefully the process of verbal perception.  
 
A suitable way of doing so is demonstrated in the next section. In section III, I employ a retrieved-context description of verbal recall to derive both WLEs (standard and inverse). In section IV, I test two key predictions of the theory against data from the PEERS experiment of Kahana et al. (Lohnas and Kahana, 2013; Healey and Kahana, 2016). In section V, I list five experiments designed to test further predictions of the theory. In the Conclusions, I sketch a possible interpretation of the results. 

\section{Verbal perception}
 
\subsection{Structure of Semantic Space}

Verbal perception is the conversion of words into meaning, and any theory of the phenomenon must begin by defining the space in which the mental trajectory takes place. I will refer to this as "semantic space" and will represent "semantic states" as integers belonging to a segment $X =  ( -A,A) $, with $A \in \mathbb{N}$. Measurable quantities are to be computed in the limit $A \rightarrow \infty$, and a generalization to higher dimensions will be introduced below. 
 
Since $X \subseteq \mathbb{Z}$, it follows that, between two discrete times $t_0$ and $t_0+N$, the psychological trajectory can be represented by a sequence of integers $x_0, x_1, \ldots, x_N$. As memory plays a crucial role in thought process, the laws of motion governing the trajectory will be in general highly non-markovian. 

\begin{figure}[h]
\label{fig2}
\includegraphics[width=1\textwidth]{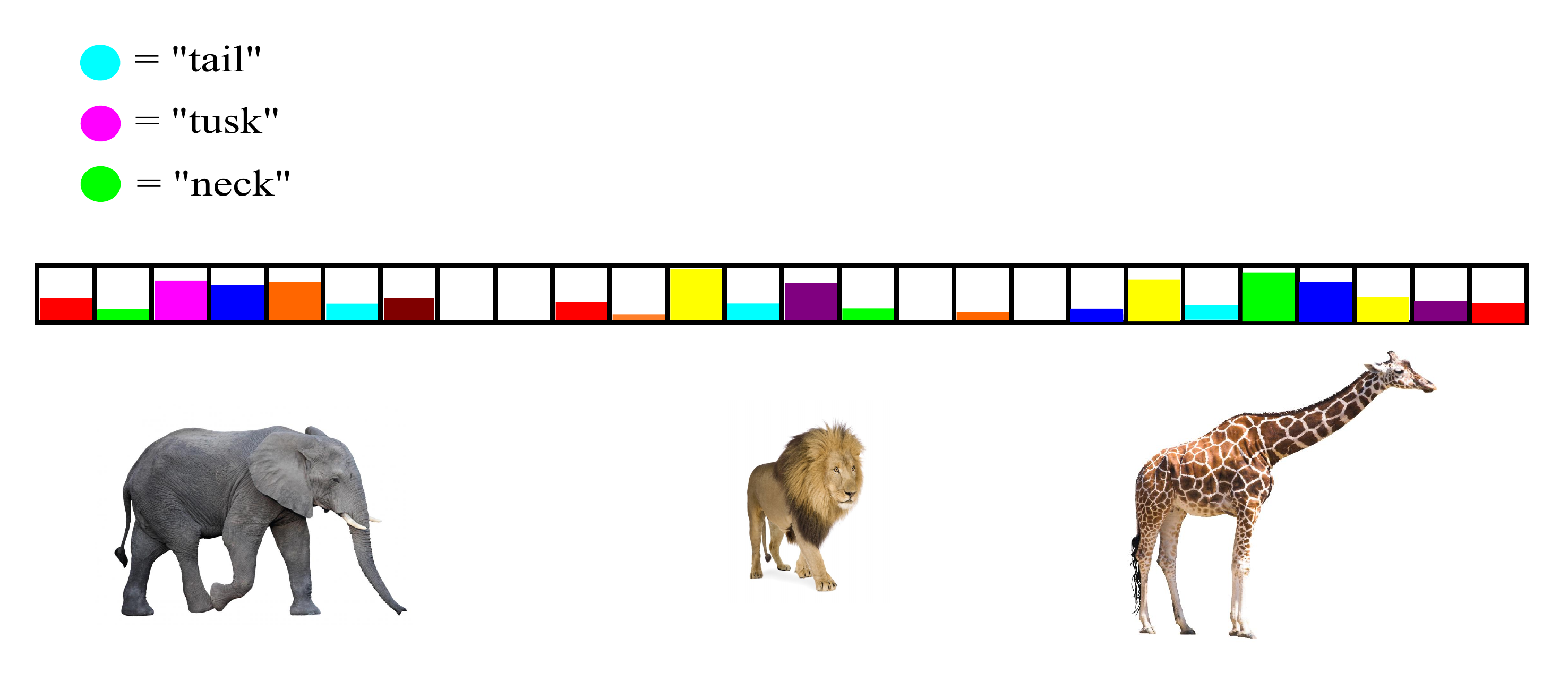}
\caption{  
Position dependence of word distribution in semantic space. The space is represented as an array of boxes. Each box represents a state, each color a word. Three of the words are typed above next to circles the color that represents them. The images underneath show meaning in different regions. States are filled with color to the extent to which they are verbalizable. The white boxes correspond to $q_x = 0$.}
\centering
\end{figure}
   
Call $\mathcal{W}$ the vocabulary available to the system. Naturally, a given word $w\in \mathcal{W}$ may be appropriate to describing more than one state, though with a varying degree of appropriateness. For each word $w$, thus, one can define its semantic range $X_w \subset X$ as the set of all states described by that word. Meanings should be seen as distributed in an universal way over $X$, as  in Figure 2. The function $w \rightarrow X_w$ matches each word with locations associated to meanings suitable for that word.

Not all states are equally fit to be verbalizable. Hence, there will be a varying probability $q_x$  that the state $x$ will match the word describing it. The value of $q_x$ is the degree of verbalizability of state $x$; or conversely, given $x \in X_w$, $q_x$ is the fitness of word $w$ as a descriptor of state $x$. 
    
If the sets $\{X_w\}_{w\in \mathcal{W}}$ do not overlap, we can define a verbalization function $v_x: \bigcup_{w\in \mathcal{W}} X_w\rightarrow  \mathcal{W}$ such that $v_x = w$ whenever $x \in X_w$. When the system tries to verbalize state $x \in X$, it produces word $v_x$ with probability $q_x$, and the silent word $0$ with probability $1 - q_x$. We will define $q_x$ so that it is only null for nonverbalizable states. 
         
Finally, we can take the trajectory to become markovian whenever it is driven by a specific verbal input: at the instant in which a new word is presented, the system hops into the nearest state described by that word.  
  
Accordingly, let the function $\xi_w : X \rightarrow X_w $ be defined by the condition: 
  
\begin{figure}[h]
\label{fig3}
\includegraphics[width=0.8\textwidth]{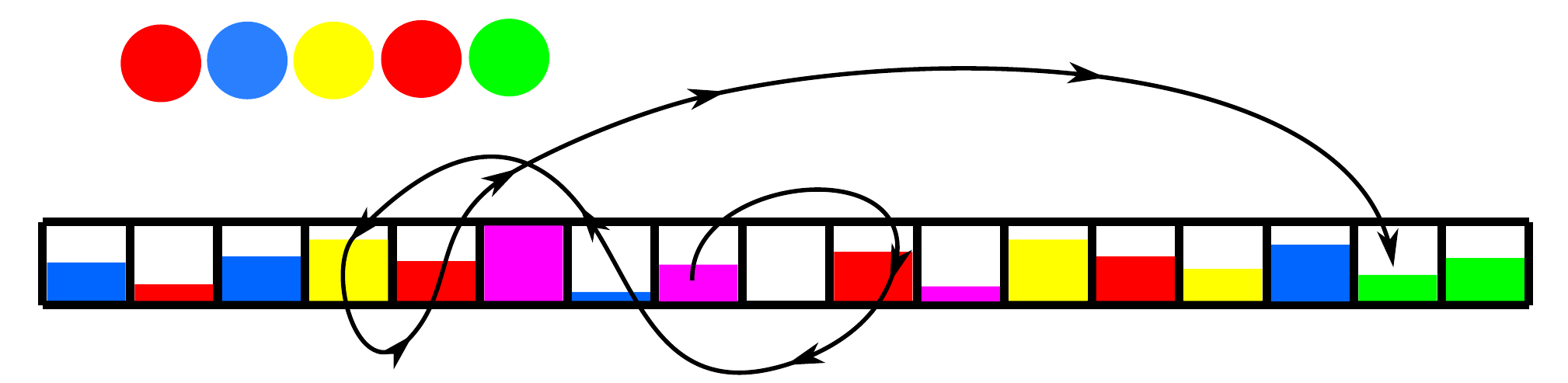}   
\caption{Representation of the trajectory induced by the perception of a verbal input. The input consists of five words, depicted here as circles. As the system perceives each new word, it reaches for the nearest state described by that word. }\centering
\end{figure} 

\begin{equation}
\label{xi}
| \xi_w (x)  - x | \leq | \tilde{x} - x|  \ \ \ \forall \tilde{x} \in X_{w}
\end{equation}

with the understanding that, if two $x\in X_w$ satisfy this condition, $\xi_w (x)$ is chosen by flipping a coin. 
  
Given a verbal input $\vec{w}= (w_1, w_2, \ldots, w_N)$, with the $w_i$'s $\in \mathcal{W}$, and a starting point $x_0$, the trajectory of the system while perceiving the input will be given by :

\begin{equation}
\label{hopping}
 x_n = \xi_{w_n} \circ  \xi_{w_{n-1}}  \circ \ldots \circ  \xi_{w_1} (x_0) \end{equation}
  
$\forall n \in (1,N)$ .
          
Obviously, the ordering of words in the input is essential. Suppose for instance that the vocabulary contains two words, $\mathcal{W}=\{A,B\}$, with $X_A=\{0,3\}$ and  $X_B = \{2\}$, and the starting point is $x_0=0$. The verbal input $(A,B)$ yields the trajectory $(0,0,2)$, while the input $(B,A)$ yields the trajectory $(0, 2, 3)$. Word $A$ corresponds to two different states in the two inputs.

$\\$

\begin{figure}[h!]
\label{fig4}
\includegraphics[width=.8\textwidth]{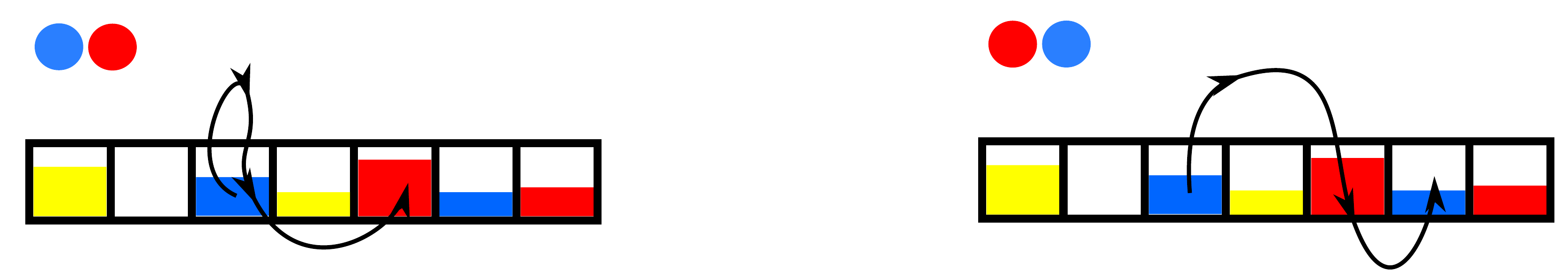}
\caption{Noncommutative nature of verbal perception. Example of trajectories induced by the permutation of two words.}\centering
\end{figure} 

\subsection{Effect of the Voronoi length distribution}
   
The verbalizable states associated to word $w$ can be seen as seeds of a Voronoi partition of $X$ into cells $\{c_x\}_{x\in X_w}$, where states located on the boundaries are understood to belong to each of two cells with probability $\frac{1}{2}$.

\begin{figure}[h]
\label{fig5}
\includegraphics[width=1\textwidth]{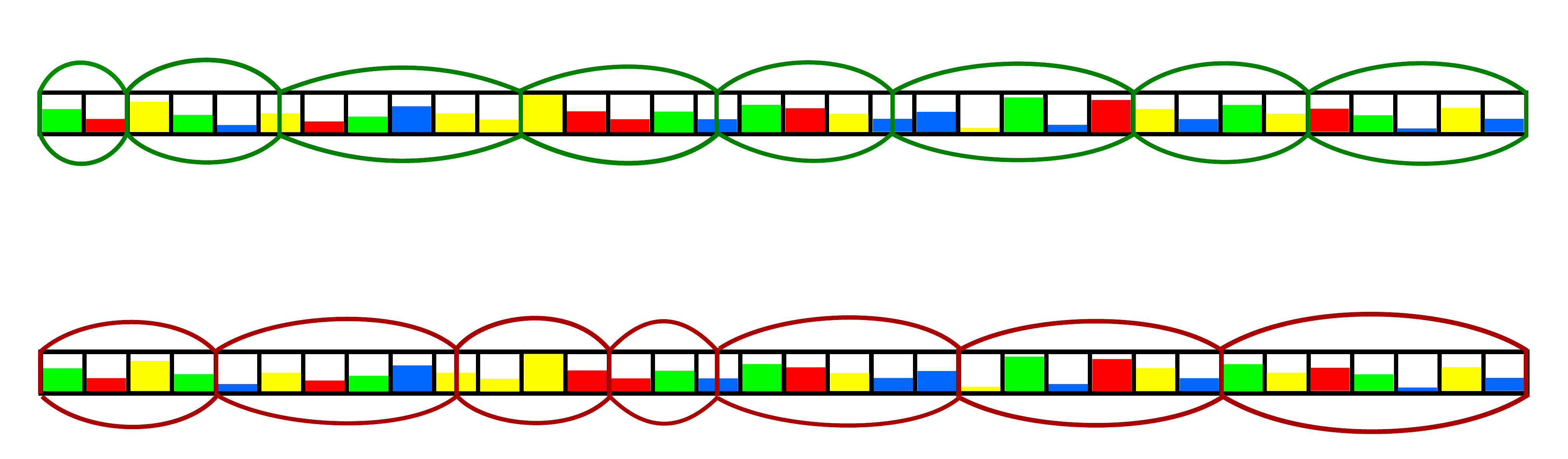}
\caption{Voronoi cells of the "red" and "green" word within the same system.}\centering
\end{figure}

During the perception of verbal inputs, the distance travelled by the system is controlled by the function $d(x,w):=
|\xi_w(x) - x|$, the "reaching distance" of word $w$ from state $x$. This is just the distance of $x$ from the closest state belonging to $X_w$. 

The level of nonlocality of a word $w$ may be measured by its reaching distance averaged over all starting points: 

\begin{equation} 
\label{reachingdistance}
d_w = \frac{1}{|X|} \sum_{x \in X} | \xi_w(x) - x| 
\end{equation}
   
Obviously $d_{w_1} > d_{w_2} $ does not necessarily imply $|X_{w_1}| > |X_{w_2}| $. Consider two words $w_1$, $w_2$ such that $X_{w_1} = \{-A, A\}$  and  $X_{w_2} = \{-A/2, A/2\}$, with $A$ even and $>2$. We have $|X_{w_1}| = |X_{w_2}| $, but $d_{w_1} = \frac{A^2}{2 A + 1} > d_{w_2} = \frac{A^2 + A}{2 (2 A + 1)}$.
  
A direct relation can be shown to exist between a word's average reaching distance and the size of its Voronoi cells. 
Call $\lambda_i$ the size of the $i$-th Voronoi cell of word $w$. 
Assuming $\bar{\lambda} \gg 1$, we may neglect boundary effects and write

\begin{equation}
\label{voronoi}
d_w \sim \frac{1}{|X|} \sum_{i \in X_w} 2 \sum_{d=1}^{\lambda_i/2}  d \sim \frac{\bar{\lambda} }{4} \Bigg( 1 + \frac{\overline{\strut (\lambda - \bar{\lambda})^2}}{\bar{\lambda}^2} \Bigg)
\end{equation}

Thus, the average reaching distance of a word depends solely on the first two moments of its Voronoi length distribution.

If $\overline{\strut (\lambda - \bar{\lambda})^2} \gtrsim \bar{\lambda^2}$ 
for a word $w$, its word structure contains strong fluctuations, so one may separate $X$ into regions were states described by $w$ are denser, and regions were they are sparser. Word $w$ is effectively "localized" inside the regions were such states are dense, that is, it expresses those semantic areas better than those where its states are sparse.

The degree of word localization has arguably a strong effect on the dynamics, as shown in Figure 6.  
 
\begin{figure}[h] 
\label{fig6}
\includegraphics[width=1\textwidth]{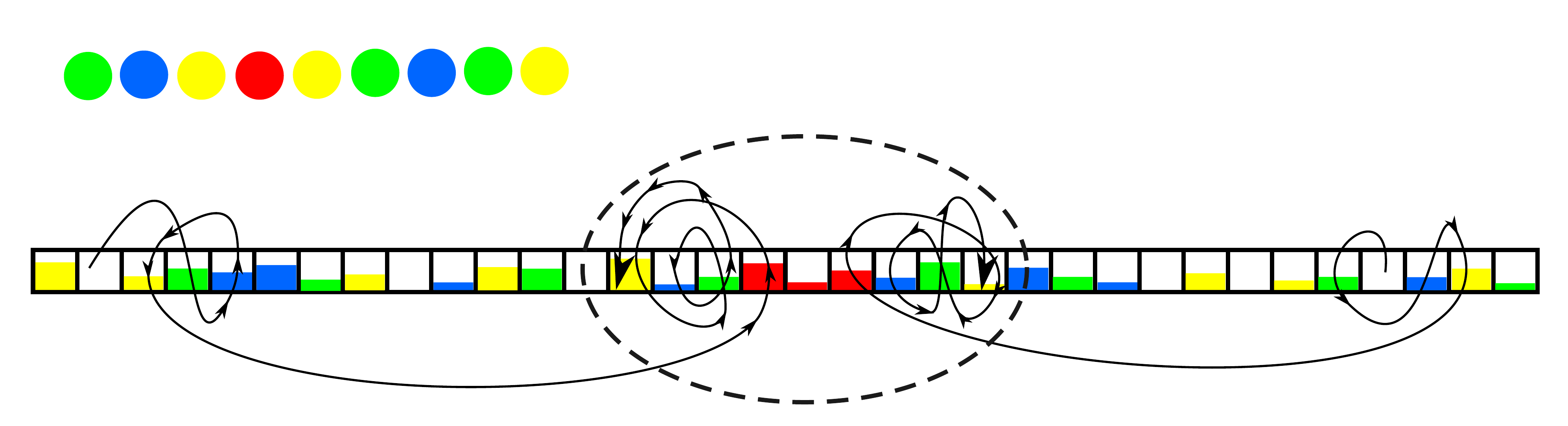}
\caption{Example of two trajectories induced by the same input (shown as an array of circles). The two trajectories begin when the system in different states. The red (localized) word creates a narrow trapping region in semantic space, marked by the dashed ellipse. Once the localized word appears in the input, all trajectories are trapped within the ellipse.}\centering
\end{figure} 
 
\subsection{Analysis of word-length related properties}  
    
In most experiments, the relevant number of word-lengths is four, since words with more than four syllables are rare in English. For simplicity, here we will consider the existence of only two word-lengths, short and long. 
  
Call $d_{\alpha}$ (with $\alpha=s,l$) the reaching distance averaged over space and over all words of the same length (short or long). From eq. (\ref{voronoi}), we see that the space-averaged reaching distance is the product of two factors, involving respectively the first and second moment of the Voronoi length distribution. Let us consider how these two moments depend on the word's length. 
    
The average Voronoi length $\bar{\lambda}$ may be related to the frequency of a given word in a corpus of the language. Indeed, if the system, in its 'speaking mode', explores semantic space ergodically and uniformly, the frequency of a word is $\nu_w = \overline{\lambda_w} \ ^{-1}$. 
  
In a typical corpus of the English language, the frequency $\nu(S)$ of words with $S$ syllable is monotonously decreasing. As a consequence, we expect the average of the Voronoi length $\overline{\lambda_w}$ to be larger for short words than for long words.  While this is correct for most languages, notice that there are exceptions, such as Turkish and Arabic, where the function $\nu(S)$ is peaked at $S = 2$ (Fucks, 1956; Grzybek, 2007).
 
Let us now look at the relative fluctuations of the Voronoi length, described by $\frac{\overline{(\lambda - \bar{\lambda})^2}}{\overline{\lambda}^2}$. We will surmise their magnitude through a qualitative argument. 
   
As mentioned in the Introduction, various approaches have been taken to prove that long words are on average more 'technical', 'specialized', 'distinctive' or 'marked' than short words. Several claims made in (Elts, 1995), (Mikk et al., 2000), and (Lewis and Frank, 2016) may be rephrased as the statement that long words are, on average, conceptually more specific. A word is conceptually specific if it is localized in certain areas of semantic space. A correlation exists, therefore, between word-length and semantic localization. 
      
Localization, in turn, will occur if the scale over which the Voronoi length fluctuates is comparable or greater than its average value. This leads to the conclusion that the relative fluctuations of the Voronoi length will be larger for longer words.  
    
Thus, both factors in eq. (\ref{voronoi}) take a greater value if the word is long. It follows that $d_l >  d_s$.

\subsection{Higher dimensions}
 
A one-dimensional modeling of semantic space is of course unrealistic, and may not suffice for every application of the theory. If $X$ is taken to be a connected subset of $\mathbb{Z}^n$, the definition we have given for the word structure $\{X_w\}_{w \in \mathcal{W}}$ applies all the same. The points in space are now vectors, and equations (\ref{hopping}) and (\ref{reachingdistance}) are still valid, the distance in eq. (\ref{reachingdistance}) being the Euclidean distance in $n$ dimensions.  
   
The Voronoi cells, however, become less simple to treat as they can be arbitrary polyhedra (for a complete treatement, see Aurenhammer et al., 2013). Formula (\ref{voronoi}) for the reaching distance must be modified, and it takes a geometry-dependent form. 
  
The Voronoi structure, yet, affects directly only the process of verbal perception, not the process of memory retrieval, which will be the subject of the next section. Thus, while in the figures I will refer to the one-dimensional case, the mathematical results will apply to any number of dimensions.
   
\section{Verbal recall} 
  
\subsection{Retrieved-context primer}
  
We have seen that the rules of motion of the system become markovian during the perception of verbal input. In retrieved-context models of short-term memory, the rules of motion are also markovian during the search for memories, which is conducted through the principle of free association. In these models, the retrieval of memories per se is not a measurable phenomenon. What can be observed is the retrieval of words describing those memories. Therefore, a recall experiment must be seen as the composition of two processes: a processes of memory retrieval, and a process of memory verbalization. 
 
The markovian process of memory retrieval must be described here, in keeping with the spirit of context-driven models, as a random walk on $X$ that effects a retrieval whenever it meets a state corresponding to the experience to be recalled. The verbalization process, on the other hand, depends on the verbalizability of memories: a memory $x$, once retrieved, has a probability $q_x$ of leading the system to produce the word describing it. 
      
\begin{figure}[h]
\label{fig7}
\includegraphics[width=0.7\textwidth]{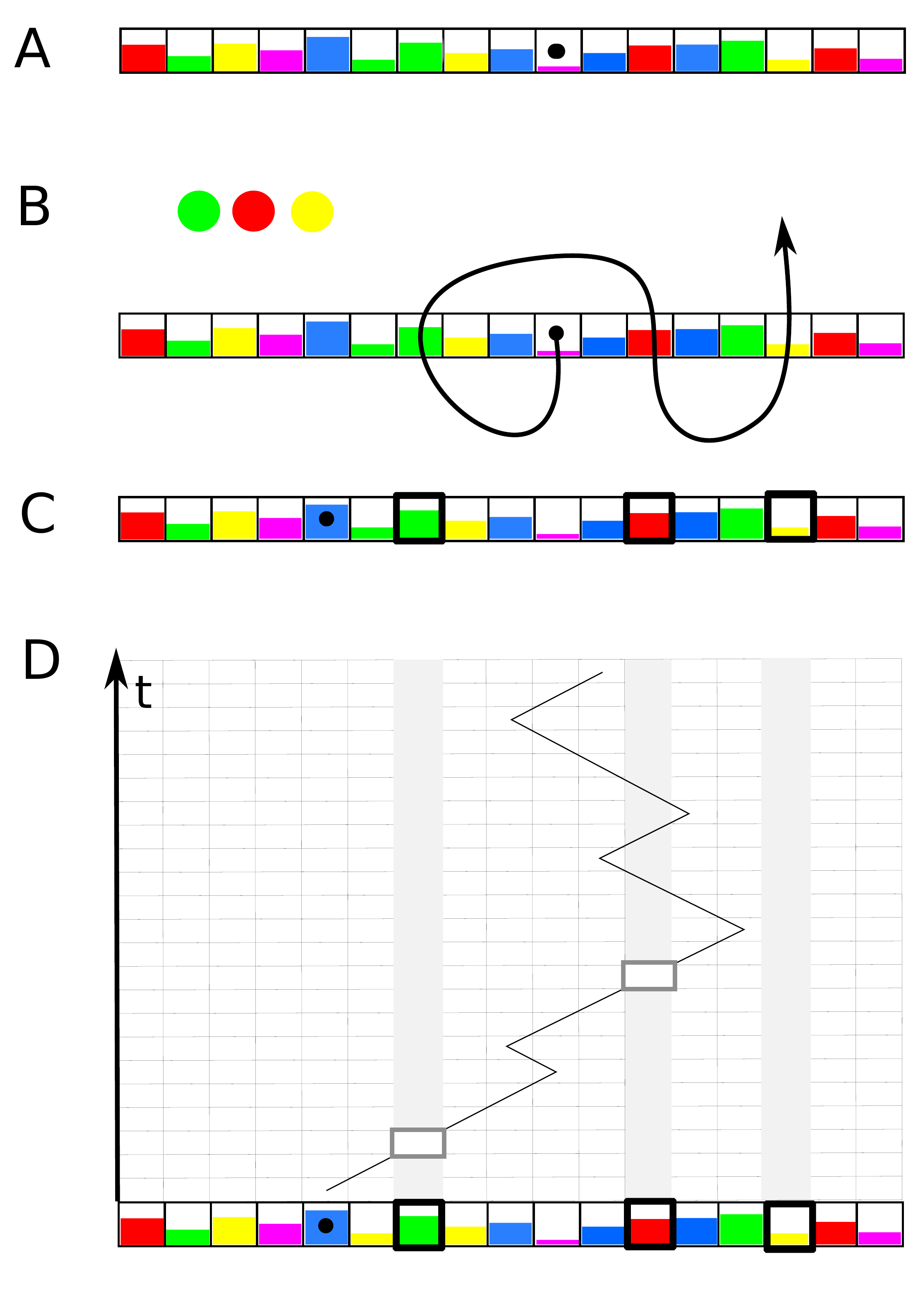}
\put(-135,420){$y_0$}
\put(-216,253){$x_0$}
\caption{Trajectory during presentation and recall of a three-word list. Starting from a random position (stage A), the system moves under the effect of verbal input (stage B), and its discontinuous path leaves memory traces (stage C) that are pursued by a random walk in the retrieval stage (stage D). The points in space-time corresponding to retrieval are boxed.}
\centering
\end{figure}

The following mathematical  problem arises. Supposing one is given
 
1) the structure $\{ v_x, q_x\}_{x}$ of the vocabulary;
   
2) a word list $\vec{w}= ( w_1, w_2, \ldots, w_N )$ presented to the system;

3) the state of the system when the word list begins to be presented, that is, a probability distribution $\chi[y_0]$ on its position $y_0$; 
 
4) the state of the system when the retrieval process begins, that is, a probability distribution $\psi[x_0]$ on the new position $x_0$; 
  
one wants to predict the probability that the $i$-th word will be among those recalled by the system. 
 
In the next section, this program will be carried out for the particular case of a vocabulary containing words of two different lengths.

\subsection{Application to the double word-length scenario}
   
We will begin by defining the probability $P_h(t)$ that a memory placed at distance $h$ from $x_0$ is met by the retrieving random walk for the first time after a time $t$. In one dimension, this is given by 
 
$$P_h(t) = \sum_{\vec{n}:  \ \sum_1^h n_i = t } f_{n_1} f_{n_2} ... f_{n_h}$$

where $f_{2 n} = 0$ and $f_{2 n -1} = \frac{(2n - 3)!!}{n!2^n}$. 

The probability that a memory will be retrieved is therefore

\begin{equation}
\label{retrieval}
p^{retrieval} (h) = \sum_0^T P_h(t)
\end{equation} 

where the cutoff $T$ is needed to obtain meaningful results in one or two dimensions, and can otherwise be let to infinity.
 
Suppose a list $\vec{w} = (w_1, w_2, \ldots , w_N)$ has been presented. Call $ p_i(\vec{w}; y_0, x_0)$ the probability of retrieving the memory created by the $i$-th word in the list, given a certain initial position $y_0$ for the trajectory during presentation, and a certain initial position $x_0$ for the trajectory during retrieval.

We have:

\begin{equation}
p_i(\vec{w}; y_0, x_0)= p^{retrieval}\bigg( \Big| \ \Xi_{i}(y_0) - x_0 \ \Big| \bigg)
\label{piv}
\end{equation}

where $\Xi_{i} := \xi_{w_i} \circ \ldots \circ \xi_{w_2} \circ \xi_{w_1}$. 
     
Our goal is to compute the average recall probability for an arbitrary list composed by $S$ short words and $L$ long words arranged into a given order. This amounts to averaging eq. (\ref{piv}) over all lists of the same type $\vec{\alpha} = (\alpha_1, \ldots, \alpha_N)$
where $\alpha_i \in \{ s,l \}$ for $i=1, \ldots, N$:

\begin{equation} 
p_i(\vec{\alpha}; y_0, x_0)= \frac{1}{W_s^S W_l^L} \sum_{\substack{\vec{w} \in \mathcal{W}^N \\ \alpha(w_i) = \alpha_i \\ i=1,\ldots, N}}
p_i(\vec{w}; y_0, x_0) := \Bigg\langle p^{retrieval}\bigg( \Big | \ \Xi_{i}(y_0) - x_0 \ \Big| \bigg)
\Bigg\rangle_{\vec{\alpha}}
\label{piv2}
\end{equation}
  
where $\alpha(w)$ is the type of word $w$ and $W_s$ ($W_l$) is the number of short (long) words in the vocabulary. 
 
We will perform the averaging over lists through a mean-field approach. Mean-field approaches, widely employed in physics, consist in inverting the order of the two steps: the computation of observables and the averaging. Instead of averaging the final probabilities, one averages an intermediate, non-observable quantity usually called the "field". In this case, we may average the functions $\xi_w$ themselves. 
   
From section IIC, we know that the average displacement induced by the function  $\xi_w$ is equal to $d_{\alpha(w)}$ with $d_s < d_l$, while no constraints emerged concerning the direction of this displacement as a function of word type. Therefore, to average the functions $\Xi_i$ over all word lists of one type, we replace $\xi_w$ with the function $\bar{\xi}_{\alpha(w)}$ defined by $\bar{\xi}_\alpha (x) = x + d_{\alpha} \hat{e}$, where $\hat{e}$ is a unity vector randomly chosen from a suitable distribution $\Omega_0[\hat{e}]$.  
   
We can thus write $\Xi_i(y_0) \sim y_0 + \sum_{k=1}^j d_{\alpha_k} \hat{e}_k $, and eq. (\ref{piv2}) becomes

\begin{equation}
\label{meanfield} 
p_{i}(\vec{\alpha}; y_0, x_0) = \Bigg\langle p^{retrieval}\Big(\Big|\sum_{k=1}^i d_{\alpha_k} \hat{e}_k  + y_0  - x_0 \Big| \Big) \Bigg\rangle_{\Omega}
\end{equation}

where $\Omega (\hat{e}_1, \ldots, \hat{e}_i) = \prod_{k=1}^i \Omega_0 [\hat{e_k} ]$. 
 
The initial positions $y_0$ and $x_0$ are not measurable quantities. In eq. (\ref{meanfield}), therefore, they must be averaged over through two suitable distributions $\chi(y_0)$ and $\psi(x_0)$, yielding  
  
\begin{equation}
\label{meanfield2}   
p_{i}(\vec{\alpha}) = \bigg \langle p^{retrieval}\Big(\Big|\sum_{k=1}^i d_{\alpha_k} \hat{e}_k  + y_0 - x_0 \Big| \Big) \bigg \rangle_{\chi, \psi, \Omega} = \bigg \langle 
p(y_i| x_0)  
\bigg \rangle_{\chi, \psi, \Omega}
\end{equation}
  
where $y_i : = \sum_{k=1}^i d_{\alpha_k} \hat{e}_k  $ and $p(y| x) : = p^{retrieval} \big(\big|y - x \big| \big) $.

The function we need to average may be rewritten as
 
\begin{equation}
\label{compound} 
p(y_i| x_0) = \mathbb{P}\big[i_1=i\big] + \sum_{j \neq i} \mathbb{P}\big[i_1=j\big] \  p(y_i | y_j)
\end{equation}
 
where $\mathbb{P}\big[i_1=i\big]$  is the probability that the $i$-th memory, $y_i$, will be the first one to be found, and $\sum_{i=1}^N \mathbb{P}\big[i_1=i\big] < 1$. Substituting eq. (\ref{compound}) into eq. (\ref{meanfield2}), we find
  
\begin{equation}
\label{meanfield4} 
p_{i}(\vec{\alpha}) = \Bigg\langle \ \big\langle \mathbb{P}\big[i_1=i\big] \big\rangle_{\chi, \psi} + \sum_{j \neq i} \big \langle \mathbb{P}\big[i_1=j\big] \big\rangle_{\chi, \psi} \ p(y_i | y_j) \Bigg \rangle_{\Omega}
\end{equation}
     
The distribution $\psi$ refers to the state $x_0$ of the system after the so-called retention interval, during which the subject freely elaborates the information gathered during presentation. A full understanding of such elaboration would require 
modeling the free motion of this system, but we have only been able to markovianize the equations of motion during a progressive searching task or in the presence of a driving input. Thus,
a reasonable ansatz is necessary. Neglecting recency effects, we can suppose $\psi$ to contain $N$ similar peaks at the locations $y_1, y_2, \ldots, y_N$ explored during presentation.  In this picture, the dependence of $\big\langle \mathbb{P}[i_1=i] \big\rangle_{\chi, \psi}$ on the index $i$ will be negligeable, so we can approximate $\Big\langle \mathbb{P}[i_1=i] \Big\rangle_{\chi, \psi}$ with a constant value $ p_0$.

Substituting this into eq. (\ref{meanfield4}), and rewriting the argument of $p^{retrieval}$, we find
 
\begin{equation}
\label{meanfield5} 
p_{i}(\vec{\alpha}) = p_0 \Bigg[ 1+ \sum_{j \neq  i}  
\Bigg \langle p^{retrieval} \Big(\Big|\sum_{k=\min (i,j)+1}^{\max(i,j)} d_{\alpha_k} \hat{e}_k  \Big| \Big) \Bigg \rangle_{\Omega} 
\ \Bigg] \end{equation}
  
where $\Omega$ is now the product of $|j-i|$ copies of the distribution $\Omega_0$, and the value of the summand depends solely on the number of long and short words located between word $i$ and word $j$. 
   
Using  $p^{retrieval}(0) = 1 $ to include the $j=i$ term into the sum, and 
noticing that the labels of the $\hat{e}_k$'s are interchangeable, we can finally rewrite eq. (\ref{meanfield5})
 as
 
\begin{equation}
\label{meanfield6} 
p_{i}(\vec{\alpha}) = p_0 \sum_{j }  
\pi(S_{ij}, L_{ij}) 
\end{equation}

where we have introduced the quantities
 
\begin{eqnarray}
\label{LS}
L_{ij} : = \sum_{\min(i,j) + 1 }^{\max(i,j)} \mathbbm{1}(\alpha_k = l) \hspace{31pt} \ \ \ S_{ij}  : = |i-j| - L_{ij} \\
\label{piMatrix}
\pi(m,q) : = \bigg \langle p^{retrieval} \Big( \Big| d_s \sum_{k=1}^{m} \hat{e}_k +  d_l \sum_{k= 1 }^{q} \hat{h}_k \Big| \Big) \bigg\rangle_{\Omega} 
\end{eqnarray} 
 
and the unit vectors $\hat{g}_k$, $\hat{h}_k$ 
are independently distributed according to $m+q$ copies of the distribution $\Omega_0$. 
 
For pure lists, eq. (\ref{meanfield6}) becomes 
\begin{equation}
\label{meanfield7} 
p_{i, \alpha}^{pure} := p_{i} (\alpha, \alpha, \ldots, \alpha) = p_0 \bigg[ 1 + \Big( \sum_{h=1}^{N-i} + \sum_{h=1}^{i-1} \Big) \pi\big(\delta_{\alpha s } h, \delta_{\alpha l} h\big)  \bigg]
\end{equation}

\subsection{Coexistence of word length effects}
    
As mentioned above, the probability of recalling the $i$-th word of the list is equal to the probability of retrieving the $i$-th \textit{memory}, multiplied by the factor $q_{w_i}$. In the previous section, we averaged the retrieval probability over all words of the same type; similarly, we must now average the verbalizability $q_x$, which we take to be distributed independently of the word structure $\{v_x\}_x$. Defining $q_\alpha$ as the average verbalizability of words of type $\alpha$, we obtain the full recall probability
    
\begin{equation}
\label{final}
P_i ( \alpha_1, \ldots, \alpha_N)= q_{\alpha_i} \ p_{i}( \alpha_1, \ldots, \alpha_N)
\end{equation}
    
As mentioned in the Introduction, the classical WLE is the experimental fact that pure lists made of shorter words are easier to remember. Of course such behavior may always be prevented, in principle, by making the verbalizability ratio $q_s/q_l$ sufficiently low. Yet, if the reported inversion of the WLE exists within this model, it must rely on the opposite requirement -- namely, that the verbalizability ratio is sufficiently high.   
 
The questions we are therefore supposed to answer are: first, whether there exists for this model a range of values of $q_s/q_l$ where both the classical and the inverse WLE occur; second, under what conditions on the parameters this may happen, and whether such conditions are relevant to current experiments.
  
Let us describe the typical experimental situation. In the experiments, word lists are generated by drawing words at random from a vocabulary $\mathcal{W}$. In a double-word length scenario, this vocabulary will contain $W_s$ short words and $W_s$ long ones. We may consider, therefore, an ensemble of lists of length $N$, where each word has a probability $\gamma: =\frac{W_l}{W_s + W_l}$ of being long, and a probability $1-\gamma$ of being short. 
 
The recall probability for the $i$-th word of the list can be averaged over all lists whose $i$-th word is of type $\alpha$, yielding $p_{i, \alpha}(\gamma) : = \langle p_i (\vec{\alpha}) \rangle_\gamma$. Substituting eq. (\ref{meanfield6}), this becomes:  
    
\begin{equation}
\label{random} 
p_{i, \alpha} (\gamma)= p_0 \Bigg[\sum_{\substack{S,L \geq 0 \\ 0\leq S+L \leq N -i}}  \Big\langle \pi(S, L) \Big\rangle_{\gamma}
 + \sum_{\substack{S,L \geq 0 \\ 0 \leq S+L \leq i - 1}}  \Big\langle \pi(S + \delta_{\alpha s}, L + \delta_{\alpha l}) \Big\rangle_{\gamma}\Bigg]
\end{equation}
    
In the first sum, which is the contribution from $j\geq i$ in eq. (\ref{final}), $S$ and $L$ stand for the number of short or long words in positions $k$ such that $i+1 \leq k \leq j$; in the second sum (the contribution from $j<i$) $S$ and $L$ stand for the number of 
short or long words in positions $k$ such that $j +1 \leq k \leq i$.  

In eq. (\ref{random}), the notation $\big\langle \ldots \big\rangle_{\gamma}$ has come to denote an averaging over $S$, $L$ performed for each separate value of $S+L$ and summed together. This is done through the binomial distribution
   
\begin{equation} 
\label{distrLS}
\Phi ( S, L )  = {L+S \choose L} \gamma^{L} (1 - \gamma)^{S} 
\end{equation}

Given that the total recall probability is $P_{i,\alpha}(\gamma) = q_{\alpha} p_{i, \alpha} (\gamma)$, the standard WLE effect ($P_{i, s} (0)> P_{i,l}(1)$) will occur if $\frac{q_l}{q_s} < \theta_{cl}$, where $\theta_{cl} = p_{i, s}(0) / p_{i, l}(1)$, and the inverse effect will occur if  
$P_{i, s} (\gamma) < P_{i,l} (\gamma)$, that is, if $\frac{q_l}{q_s} > \theta_{inv}$, where $\theta_{inv} = p_{i, s} (\gamma) / p_{i,l}(\gamma)$. The two effects coexist if $\theta_{inv} < \frac{q_l}{q_s}  < \theta_{cl}$
, which can only happen if  $\theta_{inv} < \theta_{cl}$. This condition can be rewritten as 
   
\begin{equation}
\label{inequality}
p_{i, l}(1)   p_{i, s} (\gamma) < p_{i, s}(0)  p_{i,l}(\gamma) 
\end{equation}
 
Now, it can be seen that $p_{i,\alpha}(\gamma)$ a decreasing function of $\gamma$, because by increasing $\gamma$ one transfers weight from the first to the second argument of $\pi$ inside both terms of eq. (\ref{random}), which reduces the value of the function $\pi$. Hence, we have $p_{i, l}(1)   < p_{i,l}(\gamma) $   and $ p_{i, s} (\gamma) < p_{i, s}(0)  $, from which it follows that the inequality  (\ref{inequality}) is identically satisfied.  
  
We conclude that, in this model, the WLE can undergo an inversion for any $\gamma \in ]0, 1[$, as sketched in Figure 1.

\subsection{Formulas for recall probabilities: slow-diffusion regime}
 
Consider a list of the type $( \alpha_1, \ldots, \alpha_N)$, where $\alpha_i \in \{s,l\}$, containing $L$ words and $S$ short ones. The trajectory during presentation is illustrated in Figure 8. The system begins from a random position $y_0$, and at each new word $w_i$ of type $\alpha_i$, its position is shifted forth by the operator $\xi_{w_i}$.  
   
The distance travelled at step $i$ is, in the mean field approach, equal to $d_{\alpha_i}$. So a memory produced by the presentation of a short word will be formed in the vicinity of the latest memory (that is, within a distance of order $d_s$) whereas a memory produced by the presentation of a long word will be formed at a longer distance (of order $d_l$) from the memory preceding it.  Consequently, memories are divided into clusters separated by a distance of order $d_l$ from each other. Each cluster spreads over a width of order $d_s$. 
  
\begin{figure}[h]
\label{fig8}
\includegraphics[width=1\textwidth]{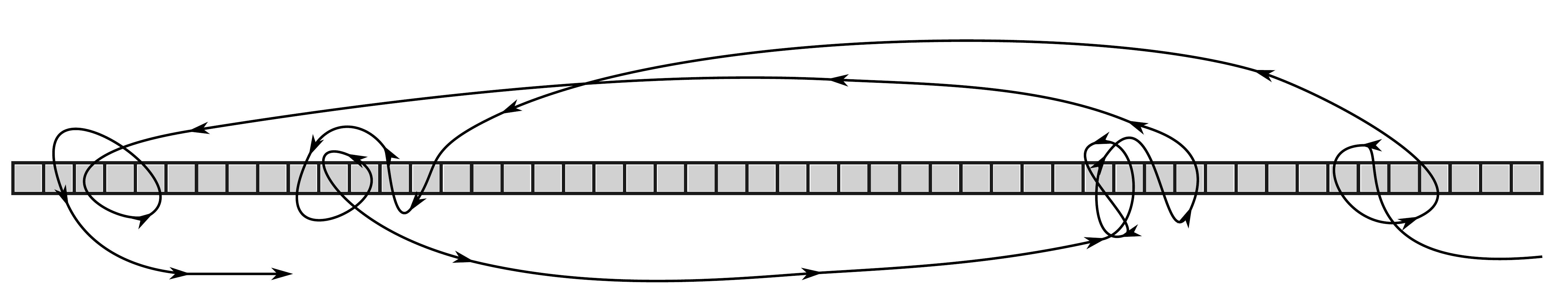} 
\caption{Trajectory during the presentation of a mixed list. The list structure corresponding to the example is $lsslsssslsssslss$, where $l$ stands for long words and $s$ for short ones. The longer jumps correspond to the presentation of long (localized) words.} 
\centering
\end{figure}
      
These memory clusters correspond to different "segments" of the list. Call $\{l_i\}_{i=1}^{L}$ the index values corresponding to long words within a given list. If $l_1=1$, the segments are $\vec{s}_i = ( w_{l_i}, w_{l_i +1}, \ldots , w_{l_{i+1}-1})$ for $i = 1, \ldots , L-1 $, and $\vec{s}_{L} = ( w_{l_{L}}, w_{l_{L} +1}, \ldots , w_{N})$. If $w_1$ is short, there is an additional segment $\vec{s}_0=(w_1, w_2, \ldots , w_{l_1-1})$, and the number of segments is $L+1$. Call $a_i$ the length of segment $\vec{s}_i$. All the $a_i$'s must be positive except $a_0$, which may be null, and $\sum_{i=0}^{L} a_i = N$. 

The direction in which the trajectory moves at each step is defined by the unknown distribution $\Omega_0$, which depends on the details of the word structure $\{v_x\}_x$. For a generic choice of $\Omega_0$, clusters formed at longer time intervals from each other will lie further apart in semantic space. Hence, the trajectory during presentation is the composition of two processes: a clustering process and a diffusion process.  
 
Notice that in eq. (\ref{meanfield5}) the argument of $p^{retrieval}$ is of the order of $\sqrt{S_{ij} d_s^2 + L_{ij} d_l^2 }$. If the list is short enough (that is, if 
$p^{retrieval} (\sqrt{ S d_s^2 + L d_l^2 }) \sim p^{retrieval} ( d_l )$, $p^{retrieval} (\sqrt{S} d_s ) \sim p^{retrieval} (d_s ) $ and $ p^{retrieval} ( d_l ) \ll p^{retrieval} (\sqrt{S} d_s ) $), 
the summand has two orders of magnitude:  one of the order of $p^{retrieval} (d_l)= p_l$ and one of the order of $p^{retrieval} (d_s)= p_s$. This corresponds to the fact that the diffusion process is slow on the scale of the trajectory during presentation. In this regime, formula (\ref{meanfield5}) for $p_i (\vec{\alpha})$ may be estimated by replacing the summand with $p_l$ whenever there is at least one long word between $\min(i,j) +1$ and $\max(i,j)$, and with $p_s$ otherwise. This is equivalent to approximating the matrix elements $\pi(m,q)$ of eq. (\ref{piMatrix}) with $\pi(1,0)$ if $q=0$ and $m>0$, and with $\pi(0,1)$ if $q>0$, thus ignoring the dependence of the retrieval process on the distance between segments of the list. 
  
Call $y_{i_1}$ the first memory to be retrieved. The conditional retrieval probability for memory $y_j$ is $p^{retrieval}(|y_j - y_{i_1}|)$. In the slow-diffusion limit, this is of the order of $p_s$ if memories $y_{i_1}$ and $y_{j}$ belong to the same cluster, and is of the order of $p_l$ otherwise. Averaging over the first retrieval, we find
 
\begin{equation}
\label{pi}
p(i) = p_0 \Big[1 + (c_i -1) p_s + (N - c_i ) p_l \Big]
\end{equation}

where $c_i$ is the length of the segment of the list to which word $i$ belongs. 
 
The average recall probabilty $P^\alpha$ for words of type $\alpha$ will thus be equal to:

\begin{eqnarray}
P_s = \frac{q_s p_0}{S} \sum_{i=0}^{L} \Big(a_i -1 + \delta_{0i} \Big) \Big[ 1 + (a_i -1) p_s + (N - a_i ) p_l \Big]\\
P_l = \frac{q_l p_0}{L} \sum_{i=1}^{L} \Big[ 1 + (a_i -1) p_s + (N - a_i ) p_l \Big]
\end{eqnarray}

Defining $\mu = \frac{N-a_0}{L}$ and $\Delta =  \frac{1}{L} \sum_{i=0}^{L} a^2_i$, we can rewrite this as 
 
\begin{eqnarray}
\label{PS}
P_s = q_s p_0\Big[ N p_l - p_s + 1 + \frac{ L}{N-L} (\Delta - \mu ) (p_s - p_l) \Big] 
\\ 
\label{PL}
P_l = q_l p_0 \Big[ N p_l - p_s + 1+ \mu (p_s - p_l) \Big]
\end{eqnarray}
 
All the dependence on the ordering of words in the list, therefore, enters the recall probabilities through the parameteres $\mu$ and $\Delta$. The values of these parameters are shown in table I for simple lists. 
   
\begin{table}[ht]
\caption{Values of the observables $\mu$, $\Delta$, and $P_s / P_l$ for simple lists}
\centering 
\label{table3}
\begin{tabular}{c c c c}
\hline \hline    
List Structure \ \ \   & \ \ \   $\mu$ \  \ & $\Delta$ & $P_s / P_l$ 
\\
\hline
\\
$l \ l \ l \ldots $
& $1$ 
& $1$ 
& \textemdash
\\ 
$ \underbrace{s \ldots s}_{S} \ l  \ldots l $ 
& $1$ 
& $1 + \frac{S^2}{N-S}$ 
& $\frac{q_s}{q_l} \frac{(N-S) p_l + (S-1) p_s + 1}{(N-1) p_l + 1}$
\\
$l \ldots l  \ s \ l \ldots  l 
$ & $\frac{N}{N-1} $
& $\frac{N+ 2}{N-1}$ 
& $ \frac{q_s}{q_l} \frac{N p_l + p_s + 1}{N^2 p_l + N - 2 N p_l + p_s - 1}$
\\ 
$\underbrace{s \ldots s}_{M}\  l \ s \ldots s  $  
& $ N-M $    
&  \ \ \ \ $N^2 - 2 M (N-M) $ \ \ \ \ 
& $ \ \frac{q_s}{q_l} \frac{N^2 p_s - 2 N p_s + p_s + N - 1 + M (2 M - 2 N + 1)(p_s- p_l ) }{(N - 1) [ (N-M) p_s + M p_l - p_s + 1] }$ 
\\
$l  \ \underbrace{s \ldots s}_{m} \ l  \ \underbrace{s \ldots s}_{m} \ldots  l \  \underbrace{s \ldots s}_{m} \  $
&\ \  $m+1$ \ \ 
& $(m+1)^2$ 
& $\frac{q_s}{q_l}$
\\ 
\\
\hline \hline
\end{tabular}
\end{table} 

For a pure list, entirely composed of words of type $\alpha$, eqs. (\ref{PS}) and (\ref{PL}) become:

\begin{eqnarray}
P_\alpha^{pure} = q_\alpha p_0 \big[ 1 + (N-1) p_\alpha \big] 
\end{eqnarray}

Data for mixed lists may be interpolated with formulas (\ref{PS}) and (\ref{PL}) to test the theory and fix the values of the internal parameters.  
  
Finally, let us look at the range of occurrence of the WLEs in the slow-diffusion regime. The classical WLE ($P_s^{pure} > P_l^{pure}$) emerges for $ \frac{q_l}{q_s} < \theta_{cl}$ where

\begin{equation}
\label{thetaClSlow}
\theta_{cl} = \frac{1 + (N-1) p_s}{1+ (N-1) p_l}
\end{equation}

The ratio $P_l/P_s$ for mixed lists can be estimated as follows. Given a fixed value of $L<N$, $\mu$ 
ranges between $\mu_{min}=1$ (for $a_0 = N-L$) 
and $\mu_{max}=N/L$ (for $a_0 =0$). 
The minimum value of $\Delta$ is obtained by starting the list with a short word and having $N - (L+ 1) \lceil \frac{N-L-1}{L +1} \rceil$ segments of $1 + \lceil \frac{N-L -1}{L +1} \rceil $ words and $(L + 1) ( 1 + \lceil \frac{N-L -1}{L +1} \rceil) -N $ segments of $1 + \lfloor \frac{N-L -1}{L +1} \rfloor$ words. The maximal  $\Delta$ is obtained by setting $a_0 = 0 $ and lumping all the short words into one segment: $\Delta_{max} = \frac{(N- L +1)^2 + L - 1}{L}$. 

Substituting $\Delta_{max}$ and $\mu_{min}$ in (\ref{PS}), (\ref{PL}), one obtains a strict upper bound on the ratio $P_s/P_l$: $\frac{P_s}{P_l} < \frac{q_s}{q_l} \theta_{inv}$, where
 
\begin{equation}
\label{thetaInvSlow}
\theta_{inv} = \frac{L^2 (p_s - p_l) + N L (p_l - 2 p_s) + p_s N^2 + (N- L) (p_s - 2 p_l +1) }{(N-L) [1 + (N -1 ) p_l ]}
\end{equation}

Once again, the inverse WLE ($P_s/P_l < 1$) will emerge under the sufficient condition $\frac{q_l}{q_s} > \theta_{inv}$; hence, the possibility that the classical and inverse effects may coexist requires $\theta_{cl} > \theta_{inv}$. 
      
If we rewrite this inequality by means of eq. (\ref{thetaClSlow}) and (\ref{thetaInvSlow}), the "microscopic" probability parameters $p_s$ and $p_l$ cancel out and we obtain 
the general condition $L^2 - L N + 2 (N - L ) < 0$, which is identically satisfied for any $L >2$. 
     
It follows that, in the slow-diffusion regime, the system displays an inversion of the WLE for \textit{all} mixed lists containing more than two long words. This statement, unlike the conclusions of the previous section, is not only true on average, but holds true regardless of the order in which short and long words are arranged within the list. 
    
\subsection{Formulas for recall probabilities: fast-diffusion regime}
     
In the previous section we have considered the case where the diffusion process was much slower than the clustering process -- which translates into upper bounds on the order of magnitude of $S$ and $L$. Notice that we defined these bounds in terms of the parameters $d_s$ and $d_l$, whose value may vary from subject to subject. Therefore, the very same list may be experienced in a fairly stationary regime by one subject, and in a regime of fast diffusion by a more easily distracted one. 
  
Here, we will consider the opposite case of very fast diffusion, defined as the regime where $p^{retrieval}(d_s + x)$ goes down fast on the length scale of $d_s$, 
 so the matrix elements of $\hat{\pi}$ decay quickly as the value of the indices grows,  
and the recall dynamics is dominated by the contiguity effect. 
If a long word causes a lesser diffusion than two short words, we may neglect all but the top-left elements of the $\hat{\pi}$-matrix: $\pi(0,0)=1$, $\pi(1,0) = p_s$, and $\pi(0,1) = p_l$. Otherwise, we can work in the lowest approximation by neglecting also $p_l$. 

Eq. (\ref{random}) becomes
 
\begin{equation}
\label{fast}
P_{\alpha} (\gamma) \sim p_0 q_{\alpha} \Big[ 1 + p_s + p_{\alpha} - (p_s - p_l) \gamma \Big]
\end{equation}
  
where I dropped the pedix $i$ because all dependence on $i$ vanishes as long as $i$ is neither $1$ nor $N$. 

The thresholds for the verbalizability ratio are 

\begin{eqnarray} 
\theta_{cl} = \frac{1 + 2 p_s}{1 + 2 p_l} \ \ \ \ \ \ \ \ \ \ \ \ \ \ \ \ \ 
\theta_{inv} = \frac{1 + 2 p_s + (p_l - p_s) \gamma}{1 + p_s + p_l + (p_l - p_s) \gamma} 
\end{eqnarray}
 
and for $\theta_{inv} < \frac{q_l}{q_s} < \theta_{cl}$, eq. (\ref{fast}) is nothing but a linear version of Figure 1. 
 
When experiments yield a near-linear curve $P_{\alpha} (\gamma)$, thus, it may be taken as a sign that the system is operating in a very fast diffusion mode.  When the observed curve is nonlinear, one must infer that terms of the type $\pi(h,0)$ for $h>2$ are relevant, and  a polynomial interpolation can be performed, by truncating the sums in eq. (\ref{random}). 
 
The crossover toward linearity of the curves $P_{\alpha} (\gamma)$ may be explored by tuning experimentally the amount of diffusion in the system, as we will see in section V.

\section{Data Analysis}

\subsection{The Data}

The foregoing analysis has proven that, within this model of verbal perception, the recall process can display consistently both the classical and the inverse WLE, as observed in the experiments.  
   
There may be, however, other mechanisms leading to a similar prediction. Katkov et al, for instance, propose a lexical explanation for both the direct and inverse WLE, based on possible differences in the long-term neural representation of long and short words (Katkov et al., 2014).  In order to tease out which mechanism is really responsible for the effects, one needs to extract further information from the data, going beyond the computation of mere recall probabilities. 

I will do so by using data from PEERS 
(Penn Electrophysiology of Encoding and
Retrieval Study), a large study conducted at the University of Pennsylania and devoted to assembling a database on the electrophysiological
correlates of memory (Lohnas and Kahana, 2013).

The sample I have considered corresponds to 
Experiment I of PEERS. It includes data from trials on $156$ college students (age range: 18$-$30) and on $38$ older adults (age range:
61$-$85 years). In each trial, $16$ words were presented one at a time on a computer screen. Each word was drawn from a pool of $1638$ words with length varying between $S=1$ and $S=6$ syllables. The word-length distribution was peaked at $S=2$, reflecting that of a typical English lexicon (but not of a typical English corpus). 

Each item was kept on the screen for $3000$ ms,
followed by an interstimulus interval of
800$-$1200 ms. After the last item in the list, there was a delay of 1200$-$
1400 ms, after which the participant was given $75$ s to attempt to recall aloud any of the just-presented items. Multiple trials were performed on each subject, summing up to $3744$ trials for the students sample and to $912$ for older adults. For more details on the experimental procedure, see Healey and Kahana, 2016. 

Because the word lengths involved are more than two, the formulas we derived in the previous section may not be applied verbatim. Nonetheless, two key consequences of the theory afford a direct comparison with the data. 

\subsection{Recall by contiguity}

We say that a word is recalled "by contiguity" when its recall occurs immediately after the recall of a word contiguous to it within the list.      
In the model we have developed, no matter how high the amount of diffusion, recall by contiguity will be more frequent for short words than for long words. This follows from the fact that two consecutive short words belong necessarily to the same segment, whereas a short word and a long word, though contiguous, may belong to different segments, and two long words are sure to belong to different segments. 
 
While this prediction has been derived in a two-length model, it generalizes immediately to models with multiple lengths. We have proven in section II that shorter words are characterized by a shorter reaching distance; hence, memories formed by shorter words have a higher chance of being located in the proximity of memories formed by contiguous words. Contiguity will play therefore a stronger role in the recall of a shorter word:  the shorter the words involved, the higher the chances of recall by contiguity. 

\begin{figure}[h]
\label{contiguity}
\includegraphics[width=.8\textwidth]{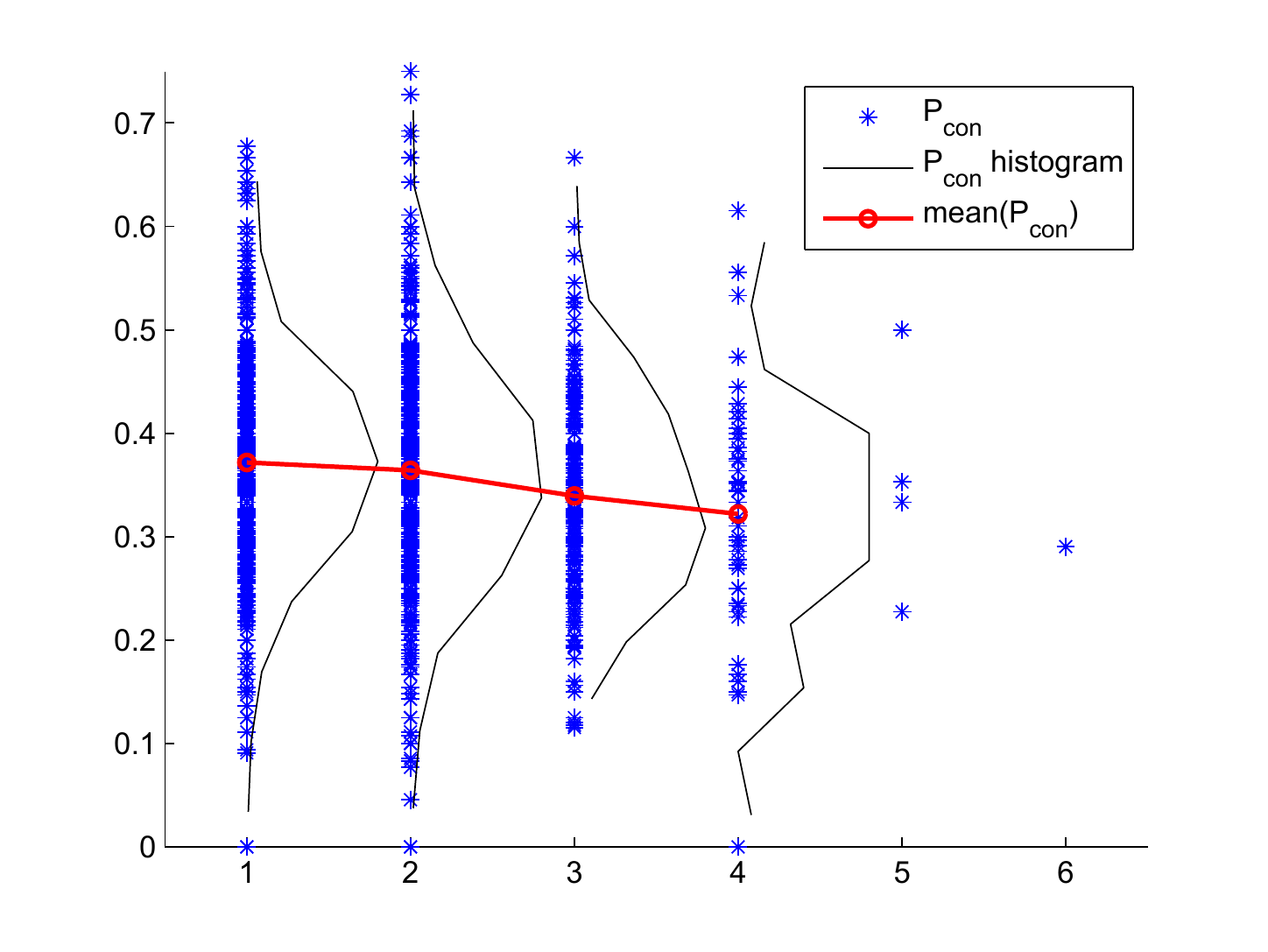} 
\put(-370,255){$P_{con}$}
\put(-47,5){ $S$}
\caption{Probability of recall by contiguity for words with $S$ syllables (blue dots); probability of recall by contiguity (black); and mean probability of recall by contiguity (red). All values have been computed from PEERS data.} 
\centering
\end{figure}

Let us call $P_{con}(w)$ the probability that a given word $w$, if recalled, will be recalled by contiguity. The theory predicts that $P_{con}$  should be larger for shorter words. If so, computing $P_{con}$ from the data and averaging it over all words of the same length would yield a decreasing curve $P_{con}(S)$. 

In Figure 9, $P_{con}$ is shown for all words having the same number of syllables (blue dots). The data have been aggregated from all trials. The distribution of $P_{con}$ (black line) is wide for all word lengths; nevertheless, the mean probability of recall by contiguity (shown in red) decreases monotonically with the number of syllables, in agreement with the theory.  
 
The $1638$-element wordpool used in PEERS contains only four $4$ five-syllable words, and a single $6$-syllable word ("encyclopedia"). Hence, the statistics for these two lengths may not be considered reliable.  

While Figure 9 refers to Experiment 1 of PEERS, the robustness of the effect has been checked for by repeating the analysis on the databases from Experiments 2 and 3 (see Lohnas et al., 2015). The decreasing trend has proven invariant across databases.

\subsection{Distribution of Jumps}

Let 
$(i_1, i_2, \ldots, i_M)$ be the serial positions of the words recalled by the subject during a certain trial. At each step $n$ in the recall process, the system performs a serial-position jump of a magnitude $\delta_n = | i_n - i_{n+1}|$. 
In our terminology, longer jumps will require the exploration of a larger portion of semantic space. Therefore, the distribution of jumps should be monotonously decreasing as a function of their size (as appears indeed from the PEERS data, Figure 10). 
 
\begin{figure}[h]
\label{distribution}
\includegraphics[width=.7\textwidth]{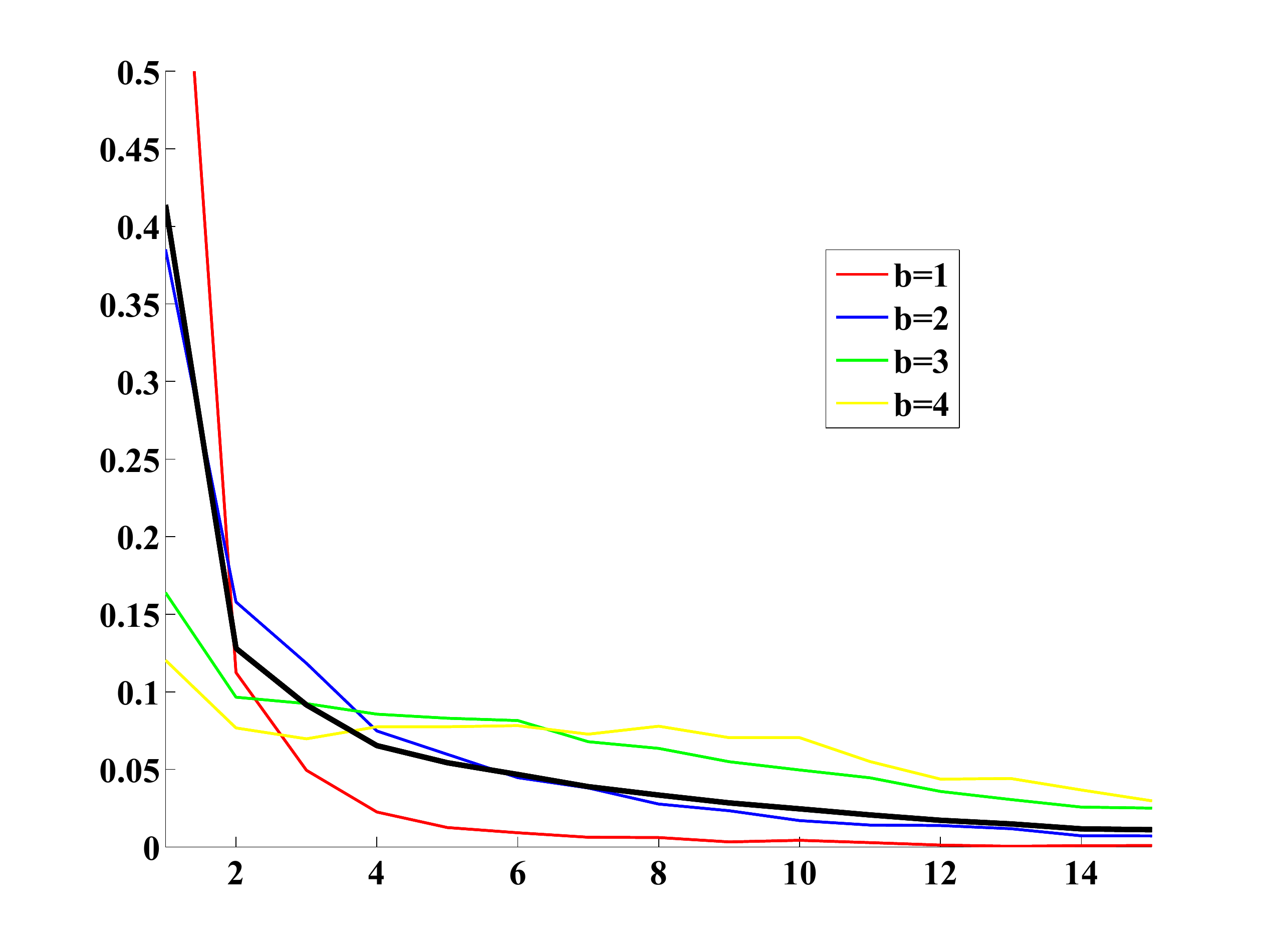} 
\put(-320,225){$P_t$}
\put(-33,5){ $\delta$}
\caption{In black, distribution of serial-position jumps in the recall process as a function of jump size $\delta$, computed from PEERS data. In colors, distribution of jumps for a fixed size $b$ of the word-length barrier, shown in the legend. As the lists contain $16$ words, $\delta$ ranges between $1$ and $15$.} 
\centering
\end{figure}
  
From our analysis of the two-length scenario, we know that these jumps cover greater distances in semantic space if the memory created by the word of departure and the memory created by the word of arrival belong to different clusters. 
Since it is long words that have the ability to break clusters, a long word located in a position $k$ such that $i < k \leq j$ plays effectively the role of a recall barrier, hindering direct transitions between $w_i$ and $w_j$. 
 
\begin{figure}[h]
\label{clusters}
\includegraphics[width=\textwidth]{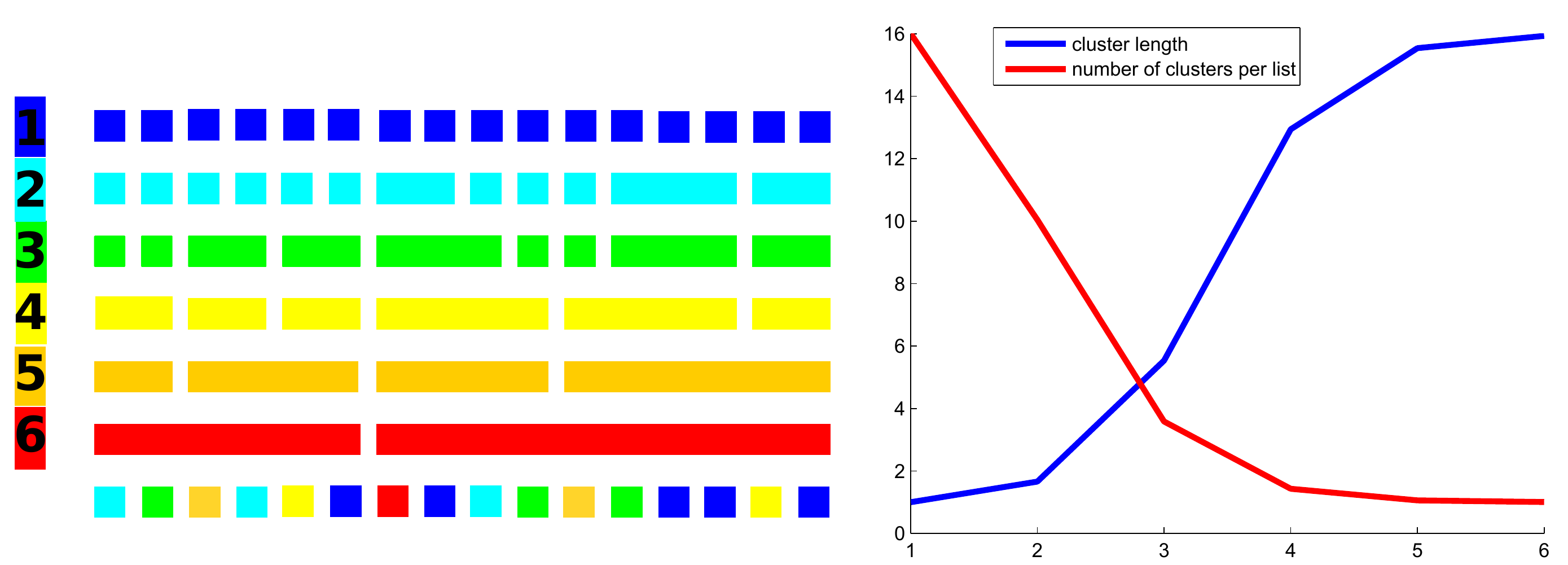} 
\put(-10,-10){ $S$}
\caption{To the left, depiction of the hyerarchical segmentation of a $16$-word list. Word lenght is encoded in colors as shown in the vertical sidebar. At each level of segmentation, clusters are displayed as horizontal bars covering the given portions of the list. To the right, a plot of the mean cluster length (blue) and of the number of clusters per list (red), as a function of the hierarchical level $S$ of clusters, computed from PEERS data by averaging over all trials. } 
\centering
\end{figure}
 
Similarly, in experiments with multiple word lengths, longer words create memories at longer distances. As clustering in semantic space will occur then over multiple length scales, clusters assume a hyerarchical structure. So does the segmentation of the list, with moderately long words marking off smaller segments within the larger segments delimited by longer words. This is shown in Figure 11, together with the average length of clusters as computed from the data, and the average number of clusters per list. 

The probability for the transition $w_i \rightarrow w_j$ (a jump of size $|i-j|$) will be affected mainly by the hyerarchical level of the clustering involved. Since longer words break clusters at a higher level, we conclude that the effect of clustering on such a jump will be controlled by the length of the longest word located between the positions $i$ and $j$: 

\begin{equation}
\label{barrier}
b_{ij} = \begin{cases} 
\max\{L(w_{j +1 }), L(w_{j +2}), \ldots, L(w_{i}) \} \ \text{if} \ \ i > j \\ 
\max\{L(w_{i +1 }), L(w_{i +2}), \ldots, L(w_{j} )\} \ \ \text{if}  \ \ j > i
\end{cases}
\end{equation}

where $L(w)$ is the length of word $w$ and, once again, the lowest index has been excluded for the same reasons why it was not counted in eq. (\ref{meanfield5}) of section III. 

\begin{figure}[h]
\label{distribution}
\includegraphics[width=.87\textwidth]{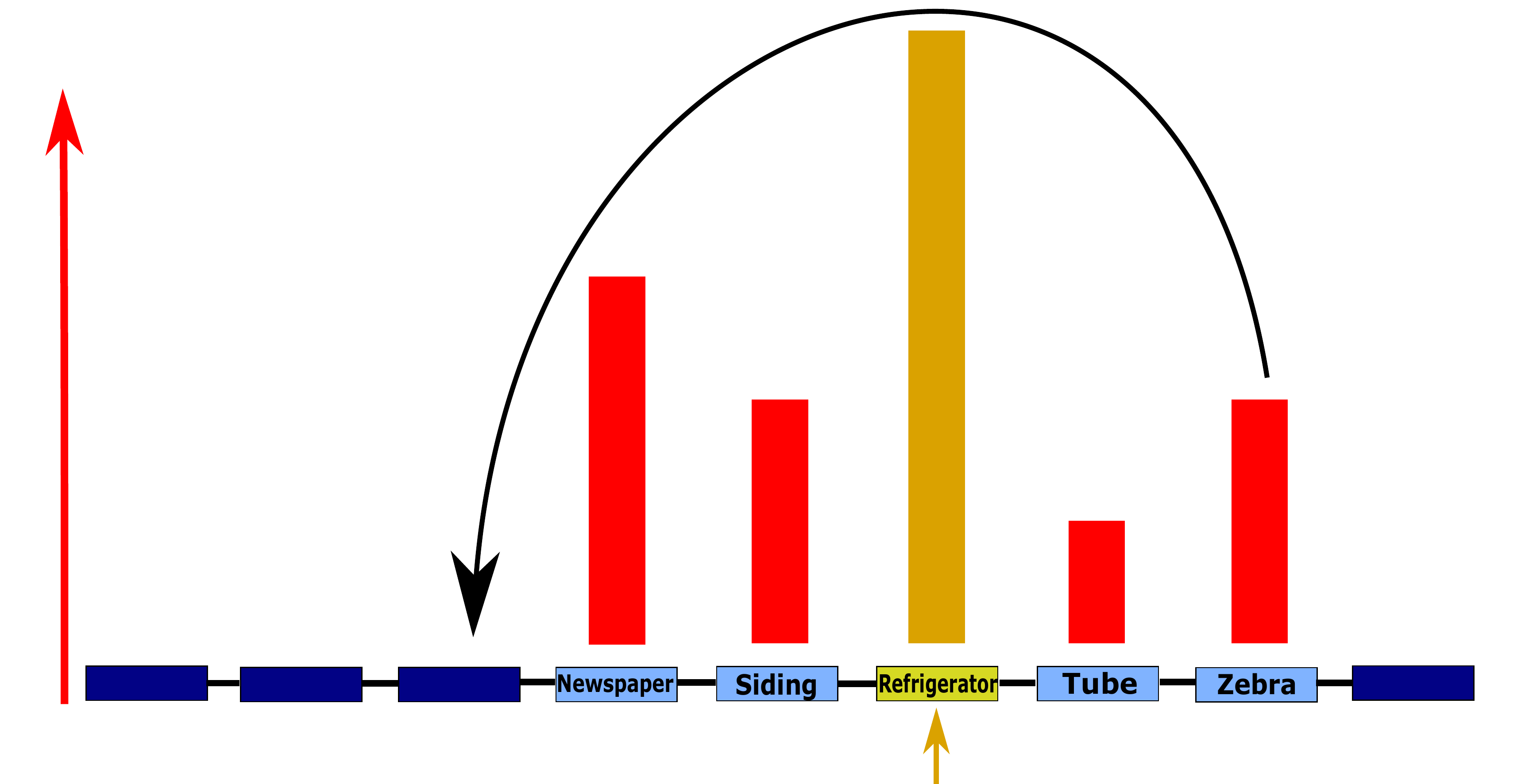} 
\caption{Serial-position jump from word $8$ to word $3$ during the recall of a nine-word list  (with words from the PEERS wordpool). The red bars describe word length: the "barrier word" for this transition is marked in yellow.} 
\centering
\end{figure}
 
We will  call $b_{ij}$ the "barrier size" for the $i\rightarrow j $ transition. The role played by this quantity is depicted in Figure 12. If the clustering mechanism does exist, it will enhance the transition probability for small values of $b_{ij}$, and suppress it for higher values. 

To test this prediction, we must first study the problem in the absence of clustering, calculating the transition probability $P_t(\delta)$ for jumps of size $\delta $. The resulting formula must be evaluated using the experimental values of parameters, and can then be compared with the values of $P_t(\delta)$ extracted from the data. If the clustering occurs, we will observe an enhancement of $P_t$ for small values of the barrier size $b$ and a suppression at high values.  
 
Let us suppose that the clustering does not occur. The system then will not "feel" the word-length barriers, and the probability $P_t$ of a jump will be independent on the size of the barrier to be overcome.  Conversely, the probability distribution of barriers over the jumps performed by the system will depend only on the jumps' size, and on the probability distribution of word length with the list. 

The probability that a jump of size $\delta$ involves a word-length barrier $b$ is equal to the probability that the longest word out of a random sequence of $\delta$ words has length $b$. This is equal to

\begin{equation}
\label{no_clustering}
P_t(b|\delta) = \left(\sum_{L=1}^b f(L) \right)^\delta - 
\left(\sum_{L=1}^{b-1} f(L)\right)^\delta 
\end{equation}

where $f(L)$ is the word-length distribution with the lists.  Formula (\ref{no_clustering}) describes the no-clustering barrier-size distribution, that is, the distribution of barrier sizes for jumps of a given length if the jumps are not influenced by the barriers.

I have evaluated formula (\ref{no_clustering}) using values of $f(L)$ computed directly from the lists recorded in the PEERS database. In Figure 13 (see the end-page), the results are compared with the values of $P_t(b|\delta)$ extracted directly from data. The no-clustering curves are dashed, the experimental curves solid. 

The effect of clustering is apparent. All experimental curves reveal an enhancement of the transition probability for small values of $b$, and a suppression at large values. Morevoer, the effect persists across the whole range of possible jump sizes. 

We would expect that the influence of clustering should be small for jumps shorter than the typical size of the clusters, and grow as the jumps get longer. This is indeed what we 
notice in Figure 13. The influence of clustering can be observed already for jumps of size one, with an enhancement of probability for small barriers and suppression for larger barriers. Yet, the clustering becomes more prominent as the jumps get longer. 

A noticeable increase in the importance of clustering is observed after $\delta=6$. This may be due to the fact that the $S=3$ clusters are becoming important, as the jumps have gotten longer than the average length of the clusters of the third level (see Figure 11). 

Since the PEERS lists have length $16$, there is no available statistics for values of $\delta$ larger than $15$. Nonetheless, we can predict that this growth will saturate when the probability of meeting a maximal bareer approaches near-certainty. For jumps longer than that, $b$ is no longer a controlling parameter; from that point on, the continued suppression of recall will derive from the need to find the next memory multiple clusters away. 

\begin{figure}[h]
\label{jumps}
\includegraphics[width=1\textwidth]{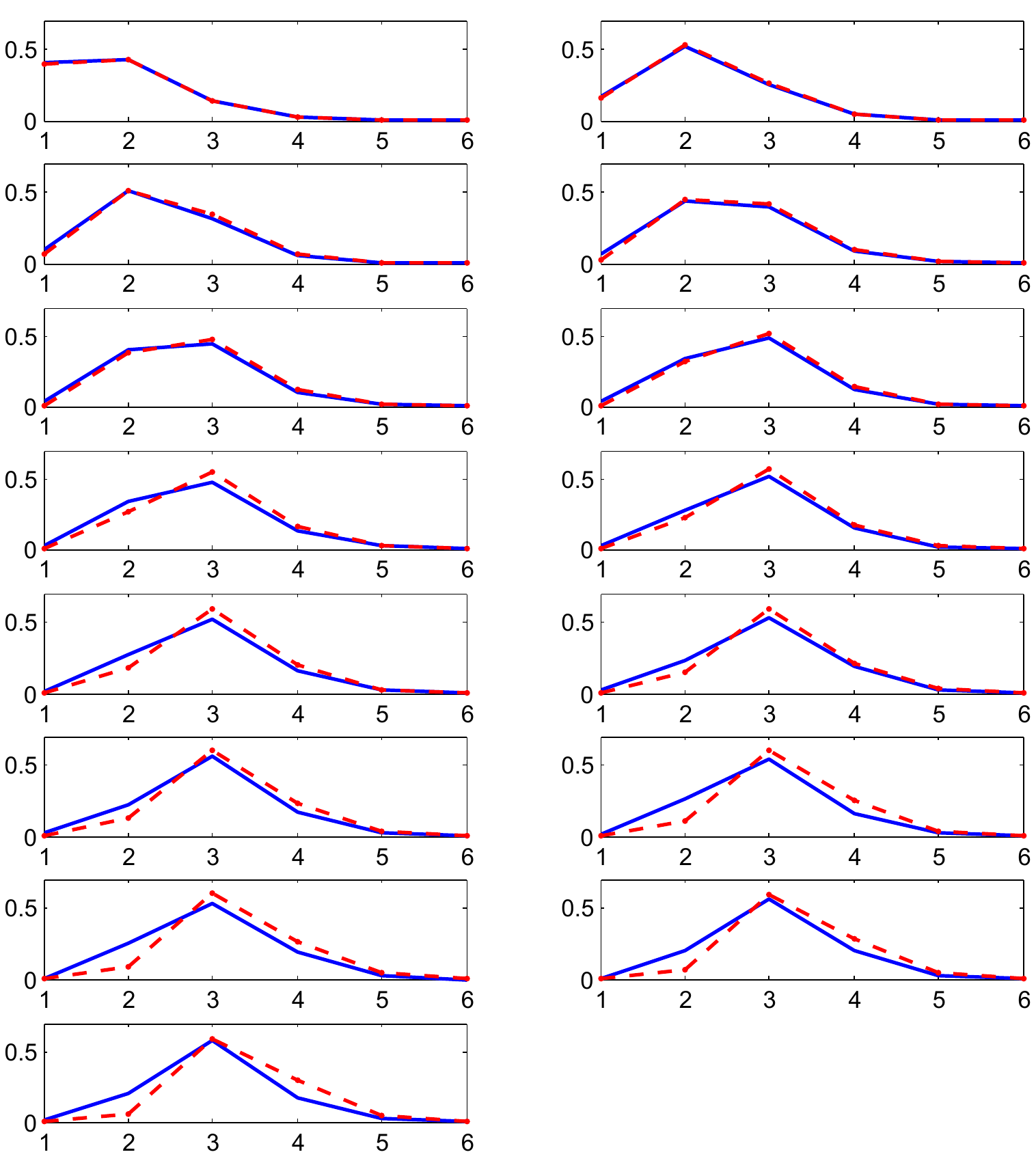} 
\put(-340,490){$\delta=1$}
\put(-340,425){$\delta=3$}
\put(-340,360){$\delta=5$}
\put(-340,295){$\delta=7$}
\put(-340,235){$\delta=9$}
\put(-340,170){$\delta=11$}
\put(-340,110){$\delta=13$}
\put(-340,43){$\delta=15$}
\put(-55,490){$\delta=2$}
\put(-55,425){$\delta=4$}
\put(-55,360){$\delta=6$}
\put(-55,295){$\delta=8$}
\put(-55,235){$\delta=10$}
\put(-55,170){$\delta=12$}
\put(-55,110){$\delta=14$}
\caption{Probability $P_t$  of serial-position jumps in the recall process, plotted as a function of the size $b$ of word-length barriers, for all possible values of jump size $\delta$. The red curves have been calculated in the absence of clustering, as per formula (\ref{no_clustering}). The blue curves are extracted from PEERS data. Clustering manifests as an enhancement of transition probability for low barriers (small $b$) and as a suppression where larger barriers are encountered. }  
\centering
\end{figure}

\section{Further Tests}

\subsection{Predictions on average recall probabililties}
 
Most free-recall experiments performed so far have employed either a wordpool with an arbitrary number of syllables, as in PEERS, or a pool characterized by a fixed length that is usually $2$ syllables (as in the Toronto Wordpool, see Friendly et al., 1982). 

Further tests on this theory, however, would be easiest to perform through experiments in which 
the words presented for recall are drawn from either of two sets: a set of very long words (e.g. number of syllables $S=4$) and a set of very short ones ($S=1$). 

When the lists are sufficiently short, the system operates in a slow-diffusion (or "clustering") regime. In this case, three parameters should be kept track of in the experiment: $\Delta$, $\mu$, and the number $L$ of long words in each list (see section III). While the recall probability of individual words depends on the details of the list, the average recall probability for short or long words should be affected by the list structure only through those three parameters, which completely characterize the list for the purposes of this type of experiment. 
    
It is then feasible to test qualitatively the following predictions:
 
1. $P^s$ correlates positively with $\Delta$ and negatively with $\mu$.

2. $P^l$ correlates positively with $\mu$.
 
These are straightforward consequences of (\ref{PS}), (\ref{PL}).  
 
In experiments with words of two lengths, data may be interpolated with formulas (\ref{PS}) and (\ref{PL}) to test the theory and to fix the values of the internal parameters. 
  
\subsection{Predictions depending on serial position}
  
Serial position effects related to word-length can be isolated experimentally by using word lists constructed on two or more basic templates of segment-structure $\vec{\alpha}=(\alpha_1, \ldots, \alpha_N)$, with $\alpha_i = s,l$. The experimenter would present equivalent lists to a number of participants and would correlate the recall probability of words with their positions within each template structure. 
 
Large data sets from such experiments should be able to corroborate or to rule out the mechanisms I have described. Competing effects of familiar types, such as primacy and recency, may be easily subtracted. 
  
Two simple predictions can be made, easier to check in a slow-diffusion regime but equally valid with faster diffusion: 
  
1. Words from longer segments have a higher recall probability than words of the same type from shorter segments;

2. Successful recall of words from one segment hinders the recall of words from others. 

The first prediction follows from eq. (\ref{pi}), and should be easy to test.  
 
The second prediction must be compared with data by computing the correlation functions $C_{ij}$, that is, the joint probability for the recall of the $i$-th and $j$-th words in the list. From there, it is straightforward to obtain the in-segment and cross-segment correlation functions:
  
\begin{eqnarray} 
C^{in}_{\alpha_1 \alpha_2} (d) = \Bigg{\langle} C_{ij} \Bigg{\rangle}_{\substack{|i -  j| = d \\ i, j \in \ \text{same segment} \\ \alpha(w_i) = \alpha_1, \alpha(w_j) = \alpha_2 }}  \hspace{40pt} C^{cross}_{\alpha_1 \alpha_2} (d) = \Bigg{\langle} C_{ij} \Bigg{\rangle}_{\substack{|i- j|=d \\ i,j \in \ \text{different segments} \\ \alpha(w_i) = \alpha_1, \alpha(w_j) = \alpha_2 }} 
\end{eqnarray}

and to verify whether $C^{in}_{\alpha_1 \alpha_2} (d) > C^{cross}_{\alpha_1 \alpha_2} (d)$, as the theory suggests.

We may add to these predictions a third one -- namely, the fact that the long word of each segment is the easiest one to recall. This results directly from the bounds we derived in the previous section on the ratio $q_s/q_l$.

\subsection{Inter-response intervals}
 
A further prediction of the theory regards inter-response intervals -- that, is the time elapsing between one recalled item and the next -- whose measurement in free-recall experiments dates back to (Murdock and Okada, 1970). 
 
During retrieval from long-term memory, it was shown (Gruenewald and Lockhead, 1980) that clusters occur due to stable semantic associations between objects: a subject who is asked to list some animals, for instance, may recall first a set of farm animals, then a number of house pets, then several birds. The inter-response intervals are shorter within clusters than between clusters. 
  
The retrieval of examples in the experiment of Gruenewald and Lockhead depends entirely on the long-term representation of items. In the situation we have described, on the contrary, short-term memory is at play, and the retrieval process has to locate the vanishing traces of a recent experience. 

Nonetheless, it is easy to see that the time interval elapsing between the retrieval of two memories will be longer between two memories belonging to different clusters, and shorter for memories belonging to the same cluster. 

Hence, the same is true among items of the list that are successfully verbalized. The inter-response intervals will be longer for the consecutive recall of two words belonging to different segments of the list, and shorter for the consecutive retrieval of two words belonging to the same segment.

\subsection{Experiments with varying presentation rate}

As mentioned in the Introduction, the WLE has been reported in experiments where the time interval between the presentation of consecutive items was a controlled parameter. Experiments on the WLE with rapid presentation of the stimuli were first performed by Coltheart and Langdon (1998), who found the WLE by presenting an item every 114 ms, every 157 ms, and every 243 ms. In (Campoy, 2008), somewhat lower presentation rates were used (between 300 and 400 ms), and again the persistence of the effect was proven over different rehearsal times.  

Here, I will argue that such experiments may offer an ideal tool to study the crossover between the "diffusive" and the "clustering" regime. 
  
Indeed, if we modify the foregoing computation to allow the system to random-walk on its own for a time $\tau$ between the presentation of the $i$-th and $i+1$-th items, this will be equivalent to increasing the average distance $d$ travelled between the memory $y_i$ generated by word $w_i$ and the memory $y_{i+1}$ generated by word $y_{i+1}$. This amounts to rescaling time while replacing $d_l$ and $d_s$  with larger effective distances. The matrix elements of $\hat{\pi}$ will decay faster as their indices grow. Therefore, the system will move closer to the diffusive regime. 

On the other hand, the theory predicts (section IIIE) that the curve $P_{\alpha}(\gamma)$ will become linear in the fast-diffusion regime. Hence, by reducing the presentation rate, one should see the two curves in Fig. 1 becoming progressively linearized, at least up to values of $\tau$ so large that not only the clustering, but also the contiguity effect breaks down. In this limit, moreover, eq. (\ref{fast}) predicts that the curves $P_s(\gamma)$ and $P_l(\gamma)$ will become parallel.   
  
Other ways of controlling the crossover between clustering and diffusing regime (e.g. by pharmacological means, or through distractor tasks such as those of Bjork and Whitten, 1974) can be similarly applied.

\subsection{Subject-dependent variability of the classical WLE}

We have just examined some ways of testing the theory that are going to require 
ad-hoc experiments. Let us conclude by mentioning one type of test that 
may be performed on already available databases -- namely,
 data from the experiments performed so far on the classical WLE. 

It is known that different people employ different strategies in order to exploit the structure of semantic space during memorization and recall (Healey et al., 2014). When it comes to the clustering described above, we expect its strength to be 
just as dependent on the subject considered. 

On the other hand, the strength of the classical WLE is also  a function of the particular subject. The difference between the recall probabilities of  all-short lists and 
all-long lists, while being positive on average, will vary in magnitude from subject to subject. 
If the mechanism at 
the root of the of the WLE is indeed a clustering phenomenon,  
we expect that it will be stronger for subjects for whom the clustering is stronger. 

Let 
$(i_1, i_2, \ldots, i_M)$ be again the serial positions of the words that have been recalled by the subject during a certain trial. The average serial-position jump 
between consecutively recalled words for this trial is $\delta_{trial} = \frac{1}{M-1} \sum_{n=1}^{M-1}| i_n - i_{n+1}|$. For each subject, we can define $\delta_{subject}$ as the average of $\delta_{trial}$ over all trials performed on him or her. The 
parameter $\delta_{subject}$ may serve as a simple measure of the subject's inclination toward clustering. Indeed, $\delta_{subject}$  is close to unity for subjects strongly inclined toward clustering, and $\delta_{subject} \gg 1$ if the subject's tendency toward clustering is low.
 
One way of gaining insight into the likelihood of the theory  consists in correlating $\delta_{subject}$ with 
the strength of the classical WLE effects, controlled by $p_s - p_L$, where $p_s$ and $p_L$ are the recall probabilities calculated 
for pure lists of short and long words. A negative correlation between these two variables 
would be a strong confirmation that the WLE is indeed due to a clustering phenomenon.
 
\section{Interpretation and conclusions}
   
I have proposed a theory of verbal perception, extracted some of its properties, and validated it through a comparison with free-recall data. 

The theory is based on the notion that a word does not  have, in general, a single meaning. A human subject exposed to a stream of verbal input will decide on the meaning of each new word on the basis of both the structure of its vocabulary and the meaning he/she has given to the words preceding it. This also applies to a list of random words, because our mind strives to interpret them as parts of a meaningful discourse. 

It may be instructive to think of such discourses as "narratives". Common experience tells us that a two-word list is already capable of creating a strong narrative sense (e.g.: picnic, lightning). When a word in the list has no semantic connection to the context created by the words preceding it, the mind perceives a "change of scenery" and assumes that a \textit{new} narrative is beginning. 

A list of words is thus perceived as a collection of distinct "stories". When prompted to recall the list, the subject remembers each story as a separate experience, and needs to re-experiences a given story before retrieving the words responsible for creating it. 
   
Words that have specific meanings have obviously less probability of fitting into a randomly generated story. Otherwise said, the words most likely to break the narrative are those with the highest level of localization in semantic space. We have argued that this correlates positively with word length. 

Hence, a list of $N$ long words is likely to break into as many one-word stories, whereas a list of short words is more likely to be perceived as a single continuous narrative. Since a single narrative is easier to recall than many unrelated ones, the standard word length effect ensues. 
    
The clustering property of short words is at play in mixed lists as well. But its effect is hindered by the presence of long words breaking the narrative. As our analysis has shown, this can lead to an inversion of the WLE, encountered in experimental observations. 
  
In this scenario, the behavior depicted by Figure 1 becomes quite logical. By replacing a short word with a long word, one splits the list into a larger number of narratives, which makes every single word in the list (whether short or long) harder to reach during the retrieval process.  
 
The interplay between the trajectory of the system during the presentation of lists and the trajectory during the memory test produces a nontrivial spectrum of behaviors, highly dependent both on the structure of lists and on the amount of "diffusion" that interferes with  clustering. 

A telltale symptom of these mechanisms is that a short word is more likely than a longer one to be recalled right after a word contiguous to it within the list. An analysis of data from the PEERS experiments (Healey and Kahana, 2016) confirms this prediction. 

Another prediction of the theory is that the serial-position jumps performed during recall will be enhanced by clustering for shorter words, and suppressed for longer ones. This is also confirmed by an analysis of PEERS data, across the whole available spectrum of jump lengths. 

Several directions stand open for experimental and theoretical work on this model. 
Experimentally, further tests may be obtained by performing the five types of measurements listed in section V. Open directions for theoretical work include: 1. generalizing the recall probability formulas to the case where the word lengths available are more than two; 2. including possible competition between words for the verbalization of a given state; 3. singling out extra effects from the fluctuations around the mean field behavior; 4. accounting for primacy and recency effects; 
5. applying the same technique to predicting the genesis of false memories.   
     
Finally, by positing a suitable mechanism for spontaneous language production, it would be useful to derive equations linking the underlying word structure to emerging verbal patterns, thus providing a direct link between the hidden variables and the observables of the model. 

I am grateful to Michael J. Kahana and his University of Pennsylvania
research group for making their raw data publicly available. I would like to thank heartily people at the Neurotheory Center of Columbia University (Ken Miller, Misha Tsodyks) for hosting me there in the fall of 2015 and for responding with enthusiasm to this theory. Thanks also to Yashar Ahmadian of the University 
of Oregon, for reading an early draft of the paper and suggesting several improvements. 

\section{Bibliography}
 	 
Aurenhammer F., Klein R., and Lee D.T., 2013. Voronoi Diagrams and Delaunay Triangulations. Singapore: World Scientific Publishing Company.

Baddeley, A.D., Hitch, G., 1974. Working memory. In G.H. Bower (Ed.), The psychology of learning and motivation: Advances in research and theory, Vol. 8, pp. 47-89. New York: Academic Press.

Baddeley A.D., Thomson N., Buchanan M., 1975. Word length and the structure of short-term memory. Journal of Verbal Learning and Verbal Behavior, 14:575-589. 

Baddeley, A.D., 2007. Working memory, thought and action. Oxford: Oxford University Press.

Bhatarad P., Ward G., Smith J. Hayes L., 2009. Examining the relationship between free recall and immediate serial recall: similar patterns of rehearsal and similar effects of word length, presentation rate and articulatory suppression. Memory and Cognition 37: 689-713.

Binet A. and Henry. V., 1894. La memoire des mots. L'annee psychologique, Bd. I 1:1-23. 

Bjork, R. A. and Whitten, W. B., 1974. Recency-sensitive retrieval processes in long-term free recall. Cognitive Psychology, 6: 173–189.

Campoy G.,  2008. The effect of word length in short-term memory:
Is rehearsal necessary? Quarterly Journal of Experimental Psychology,  61:5, 724-734.
 
Campoy, G., 2011. Retroactive interference in short-term memory and the word-length effect. Acta Psychol.  138, 135-142. 
 
Coltheart, V., Langdon, R., 1998. Recall of short word lists presented visually at fast rates: Effects of phonological similarity and word length. Memory and Cognition, 26, 330–342.
 
Elts, J., 1995. Word length and its semantic complexity, in \textit{Family and textbooks}: 115-126. Tartu: University of Tartu. 

Friendly M., Franklin P.E., Hoffman D., Rubin D.C., 1982. The Toronto Word Pool: Norms for imagery, concreteness, orthographic variables, and grammatical usage for 1,080 words. Behavior Research Methods And Instrumentation,  Vol. 14(4), 375-399.

Fucks W., 1956. Die Mathematischen Gesetze der Bildung von Sprachelementen aus Ihren Bestandteilen,
 Nachrichtentechnische Fachberichte 3:7-21.

Greenberg, J., 1966. Universals of language. Cambridge, MA: MIT Press.

Gruenewald, P.J. and Lockhead, G.R., 1980. The free recall of category examples. J. Exp. Psychol. [Hum- Learn]. 6, 225-240.

Grzybek P., 2007. History and methodology of word-length studies, in 
"Contributions to the Science of Text and Language", Dordrecht: Springer, pp. 15-90.

Haspelmath, M., 2006. Against markedness (and what to replace it with). Journal of Linguistics, 42 (01), 25-70.

Healey M., Crutchley P., Kahana MJ., 2014. Individual differences in memory
search and their relation to intelligence. J Exp Psychol 143: 1553–1569.

Healey, M. K. and Kahana, M. J., 2016. A four-component model of age-related memory change. Psychological Review, 123(1), 23-69. 

Howard, M. W. and Kahana, M. J., 2002a. A distributed representation of temporal context. Journal of Mathematical Psychology, 46, 269-299.
 
Howard, M. W. and Kahana, M. J., 2002b. When does semantic similarity help episodic retrieval? Journal of Memory and Language, 46, 85-98. 

Hulme, C., Suprenant, A. M., Bireta, T. J., Stuart, G., and Neath, I., 2004. Abolishing the word-length effect. J. Exp. Psychol. Learn. Mem. Cogn. 30, 98-106. 

Jalbert, A., Neath, I., Bireta, T. J., and Surprenant, A. M., 2011. When does length cause the word length effect? J. Exp. Psychol. Learn. Mem. Cogn. 37, 338-353.

Kahana M. J., 1996. Associative retrieval processes in free recall. Memory  and Cognition 24:103-9.

Kahana M. J., 2012. Foundations of Human Memory, Oxford University Press.

Katkov M., Romani S., Tsodyks M., 2014. Word length effect in free recall of randomly assembled word lists. Frontiers of Computational Neuroscience 8:129.

Klare, G.R., 1988. The formative years. In B.L. Zakaluk  and 
S. J. Samules (Eds.), \textit{Readability, its past, present and future}, Newark, Delaware: IRA, pp. 14-34. 

Kruglanski A.M. and Tory Higgin E., 2007. Social Psychology: Handbook of Basic Principles, The Guilford Press; Second Edition edition (p.642).

Lohnas, L.J., Polyn S.M.,Kahana, M.J. (2015) Expanding the scope of memory search: Modeling intralist and intralist effects in free recall, Psychological Revier, Vol. 122, No. 2, 337–363.

Lohnas, L. J. and Kahana, M. J. (2013). Parametric effects of word frequency
effect in memory for mixed frequency lists. Journal of Experimental
Psychology: Learning, Memory, and Cognition, 39, 1943–1946.

Mahowald K., Fedorenko E., Piantadosi S.T., Gibson E., 2012. Info/information theory: speakers actively choose shorter words in predictable contexts, Cognition, 126: 313-318.
 
Lewis, M. L. and Frank M.C., 2016. The length of words reflects their conceptual complexity. Cognition 153: 182-195.

Mikk. J., Heli U., and Elts J., 2001. Word length as an indicator of semantic complexity, in \textit{Text as a linguistic paradigm: levels, constituents, constructs}, Festschrift in honour of Ludek Hrebıcek. Trier, 187-195.
 
Miller J.F., Weidemann C.T., Kahana M.J., 2012. Recall termination in free recall. Memory and Cognition 40: 4, Pages: 540 - 550.

Murdock B-B, 1960. The immediate retention of unrelated words. Journal of Experimental Psychology 60:222-234.

Murdock B. B., 1962. The serial position effect of free recall. Journal of Experimental
Psychology, 64(5)M:482-488.

Murdock B.B. and Okada R., 1970. Interresponse times in single-trial free recall. Journal of Experimental Psychology 86:263-267.

Neath I., Bireta T.J., Surprenant A.M., 2003. The time-based word length effect and stimulus set specificity,
Psychonomic Bulletin Rev. Jun;10(2):430-4.

Neath I., Brown G. D. A., 2006. SIMPLE: further applications of a local distinctiveness model of memory, in The Psychology of Learning and Motivation, ed. Ross B. H., editor. (San Diego, CA: Academic Press), 201-243.

Piantadosi S. T., Tily H. and Gibson E., 2011. Word lengths are optimized for efficient communication, Proceedings of the National Academy of Sciences, 108, 9:3526. 

Piantadosi S. T., Tily H. and Gibson E., 2011B. Reply to Reilly and Kean: Clarifications on word length and information content, Proceedings of the National Academy of Sciences, 108, 20: E109. 

Polyn, S. M., Norman, K. A., and Kahana, M. J., 2009a. A context maintenance and retrieval model of organizational processes in free recall. Psychological Review, 116, 129-156.

Polyn, S. M., Norman, K. A., and Kahana, M. J., 2009b. Task context and organization in free recall. Neuropsychologia, 47, 2158-2163.

Roberts W.A., 1972. Free recall of word lists varying in length and rate of presentation: a test of total-time hypotheses. Journal of Experimental Psychology 92:365-372.

Romani S., Pinkoviezky I., Rubin A., Tsodyks M., 2013. Scaling laws of associative memory retrieval. Neural Computation 25:2523-2544.

Russo R. and Grammatopoulou N., 2003. Word length and articulatory suppression affect short-term and long-term recall tasks. Memory and Cognition 31:728-737. 

Sederberg, P. B., Howard, M. W., and Kahana, M. J., 2008. A context-based theory of recency and contiguity in free recall. Psychological Review, 115(4), 893-912. 
 
Standing L., 1973. Learning 10.000 pictures. Quarterly Journal of Experimental Psychology 25:207-222.

Tehan G. and Tolan G.A., 2007. Word length effects in long-term memory. Journal of Memory and Language 56:35-48.

Xu Zhan and Li Bi-Qin, 2009. The Mechanism of Reverse Word Length Effect of Chinese in Working Memory. 
  Acta Psychologica Sinica, Vol. 41 Issue (09): 802-811.

\end{document}